\documentclass[aos,preprint]{imsart}

\usepackage{xcolor}

\usepackage[normalem]{ulem}

\RequirePackage[OT1]{fontenc}
\RequirePackage{amsthm,amsmath,amssymb,amsfonts}
\RequirePackage[numbers]{natbib}
\RequirePackage[colorlinks,citecolor=blue,urlcolor=blue]{hyperref}
\usepackage{xspace,tikz}
\usepackage{rotating,multirow}
\usepackage{pdflscape,cancel}
\usepackage{algorithm,algorithmic}
\usepackage{mathtools,nicefrac,bm,graphicx,xr}

\arxiv{arXiv:1703.06222}

\startlocaldefs
\numberwithin{equation}{section}
\theoremstyle{plain}
\newtheorem{thm}{Theorem}[section]

\newtheorem{proposition}{Proposition}
\newtheorem{lemma}{Lemma}
\theoremstyle{definition}
\newtheorem{definition}{Definition}

\newcommand{\secref}[1]{Section~\ref{sec:#1}}
\newcommand{\appref}[1]{Appendix~\ref{app:#1}}
\newcommand{\lemref}[1]{Lemma~\ref{lem:#1}}
\newcommand{\propref}[1]{Proposition~\ref{prop:#1}}
\newcommand{\thmref}[1]{Theorem~\ref{thm:#1}}
\newcommand{\eqnref}[1]{\eqref{eqn:#1}}

\def\R{\mathbb{R}}

\newcommand{\tbeta}{\widetilde\beta}
\newcommand{\defn}{\ensuremath{:\, =}}

\newcommand{\Dset}{\ensuremath{\mathcal{D}}}
\newcommand{\Rplus}{\ensuremath{[0,\infty)}}

\newcommand{\PP}[1]{\textnormal{Pr}\!\left\{{#1}\right\}} 
\newcommand{\EE}[1]{\mathbb{E}\left[{#1}\right]} 
\newcommand{\EEst}[2]{\mathbb{E}\left[{#1}\ \middle| \ {#2}\right]} 
\newcommand{\PPst}[2]{\textnormal{Pr}\!\left\{{#1}\ \middle| \ {#2}\right\}} 

\newcommand{\grp}{\textnormal{grp}}
\newcommand{\nulls}{\mathcal{H}_0}
\newcommand{\nullsg}{\mathcal{H}_0^{\grp}}
\newcommand{\nullsm}{\mathcal{H}_0^{(m)}}

\newcommand{\fdp}{\textnormal{FDP}}
\newcommand{\fdr}{\textnormal{FDR}}
\newcommand{\fdpm}{\fdp^{(m)}}
\newcommand{\fdphat}{\widehat{\fdp}}

\renewcommand{\th}{\widehat{t}}

\newcommand{\bh}{\textnormal{BH}}
\newcommand{\by}{\textnormal{BY}}

\newcommand{\simes}{\textnormal{Simes}}
\newcommand{\rsimes}{\textnormal{rSimes}}

\newcommand{\pf}{\textnormal{\texttt{p-filter}}}

\newcommand{\Sh}{\widehat{\mathcal{S}}}
\newcommand{\Shm}{\Sh^{(m)}}

\newcommand{\Kset}{\widehat{\mathcal{K}}}
\newcommand{\One}[1]{{\bf{1}}\left\{{#1}\right\}}

\newcommand{\kh}{\widehat{k}}

\newcommand{\pih}{\widehat{\pi}}

\newcommand{\ITOZERO}[1]{{#1}^{i \to 0}}
\newcommand{\tf}{\widetilde{f}}
\def\L{L}

\makeatletter
\newcommand{\dotfrac}[2]{
\mathchoice
{\ooalign{$\genfrac{}{}{0pt}{0}{#1}{#2}$\cr\leavevmode\cleaders\hb@xt@ .22em{\hss $\displaystyle\cdot$\hss}\hfill\kern\z@\cr}}
{\ooalign{$\genfrac{}{}{0pt}{1}{#1}{#2}$\cr\leavevmode\cleaders\hb@xt@ .22em{\hss $\textstyle\cdot$\hss}\hfill\kern\z@\cr}}
{\ooalign{$\genfrac{}{}{0pt}{2}{#1}{#2}$\cr\leavevmode\cleaders\hb@xt@ .22em{\hss $\scriptstyle\cdot$\hss}\hfill\kern\z@\cr}}
{\ooalign{$\genfrac{}{}{0pt}{3}{#1}{#2}$\cr\leavevmode\cleaders\hb@xt@ .22em{\hss $\scriptscriptstyle\cdot$\hss}\hfill\kern\z@\cr}}
}
\makeatother

\newenvironment{carlist}
 {\begin{list}{$\bullet$}
 {\setlength{\topsep}{0in} \setlength{\partopsep}{0in}
  \setlength{\parsep}{0in} \setlength{\itemsep}{\parskip}
  \setlength{\leftmargin}{0.08in} \setlength{\rightmargin}{0.08in}
  \setlength{\listparindent}{0in} \setlength{\labelwidth}{0.08in}
  \setlength{\labelsep}{0.1in} \setlength{\itemindent}{0.1in}}}
 {\end{list}}

\newcommand{\bcar}{\begin{carlist}}
\newcommand{\ecar}{\end{carlist}}

\endlocaldefs

\begin{document}

\begin{frontmatter}
\title{A Unified Treatment of Multiple Testing with Prior Knowledge using the p-filter}
\runtitle{Unified Multiple Testing with Prior Knowledge}

\begin{aug}



\author{\fnms{Aaditya K.} \snm{Ramdas}\thanksref{m0}\ead[label=e1]{aramdas@cmu.edu}},
\author{\fnms{Rina F.} \snm{Barber}\thanksref{m2}\ead[label=e2]{rina@uchicago.edu}},\\
\author{\fnms{Martin J.} \snm{Wainwright}\thanksref{m1}\ead[label=e3]{wainwrig@berkeley.edu}},
\and
\author{\fnms{Michael I.} \snm{Jordan}\thanksref{m1}\ead[label=e4]{jordan@stat.berkeley.edu}}
\affiliation{Carnegie Mellon University\thanksmark{m0}, University of Chicago\thanksmark{m2} \\ and University of California, Berkeley\thanksmark{m1}}

\runauthor{Ramdas, Barber, Wainwright and Jordan}

\address{Aaditya K. Ramdas\\
Department of Statistics and Data Science\\
Carnegie Mellon University\\
\printead{e1}}

\address{Rina F. Barber\\
Department of Statistics\\
University of Chicago\\
\printead{e2}}

\address{Martin J. Wainwright\\
Departments of Statistics and EECS\\
University of California, Berkeley\\
\printead{e3}}

\address{Michael I. Jordan\\
Departments of Statistics and EECS\\
University of California, Berkeley\\
\printead{e4}}

\end{aug}

\begin{abstract}
 There is a significant literature on methods for incorporating
 knowledge into multiple testing procedures so as to improve their
 power and precision.  Some common forms of prior knowledge include
 (a) beliefs about which hypotheses are null, modeled by non-uniform
 prior weights; (b) differing importances of hypotheses, modeled by
 differing penalties for false discoveries; (c) multiple arbitrary
 partitions of the hypotheses into (possibly overlapping) groups; and
 (d) knowledge of independence, positive or arbitrary dependence
 between hypotheses or groups, suggesting the use of more aggressive
 or conservative procedures.  We present a unified algorithmic
 framework called $\pf$ for global null testing and false discovery
 rate (FDR) control that allows the scientist to incorporate all four
 types of prior knowledge (a)--(d) simultaneously, recovering a
 variety of known algorithms as special cases.
\end{abstract}



\end{frontmatter}

\section{Introduction}\label{sec:intro}

Multiple hypothesis testing is both a classical and highly active
research area, dating back (at least) to an initially unpublished 1953
manuscript by Tukey entitled ``The Problem of Multiple
Comparisons''~\cite{tukey1953problem,tukey1994}. Given a large set of null
hypotheses, the goal of multiple testing is to decide which subset to
reject, while guaranteeing some notion of control on the number of
false rejections.  It is of practical importance to incorporate
different forms of prior knowledge into existing multiple testing
procedures; such prior knowledge can yield improvements in power and
precision, and can also provide more interpretable
results. Accordingly, we study methods that control the False
Discovery Rate (FDR) or test the global null (GN) hypothesis while
incorporating any number of the following strategies for incorporating
prior knowledge: (a) the use of prior weights, (b) the use of penalty
weights, (c) the partitioning of the hypotheses into groups, (d) the
incorporation of knowledge of the dependence structure within the
data, including options such as estimating and adapting to the unknown
number of nulls under independence, or reshaping rejection thresholds
to preserve error-control guarantees in the presence of arbitrary
dependence. It is a challenge to incorporate all of these forms of
structure while maintaining internal consistency (coherence and
consonance) among the pattern of rejections and acceptances, and most
existing work has managed to allow only one or two of the four
strategies (a), (b), (c), (d) to be employed simultaneously.  We
present a general unified framework, called \texttt{p-filter}, for
integrating these four strategies while performing a GN test or
controlling the FDR.  The framework is accompanied by an efficient
algorithm, with code publicly available at
\href{https://www.stat.uchicago.edu/~rina/pfilter.html}{https://www.stat.uchicago.edu/$\sim$rina/pfilter.html}.
This framework allows scientists to mix and match techniques, and use
multiple different forms of prior knowledge simultaneously. As a
by-product, our proofs often simplify and unify the analysis of
existing procedures, and generalize the conditions under which they
are known to work.


\textbf{Organization}. The rest of this paper is organized as
follows. In \secref{contributions}, we begin with an example to
provide intuition, and we discuss the contributions of this paper.  In
\secref{pf+}, we describe the general \pf~framework, along with its
associated theoretical guarantees; this section lays out the central
contribution of our work. In \secref{lemmas}, we present three lemmas
that provide valuable intuition and are central to the proof of
\thmref{pf+}; see \secref{proof_main} for the proof itself. We prove
the three aforementioned lemmas in \secref{lemmas-proofs}, and prove
some related propositions in \secref{prop-proofs}. While directly
related work is discussed immediately when referenced, we end by
overviewing other related work in \secref{disc}.

\section{An example, and our contributions}\label{sec:contributions}

\begin{figure}[h]
\centering
\includegraphics[width=0.95\textwidth]{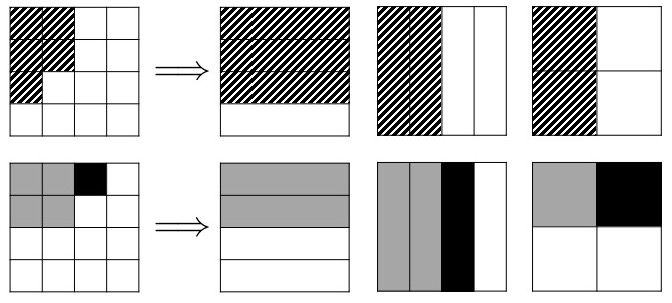}
\caption{\small Consider a set of $n = 16$ hypotheses arrayed in a $4
  \times 4$ grid, with four different partitions into groups:
  elementary, rows, columns and blocks. On the top row is the
  underlying truth, with the leftmost panel showing the
  hypothesis-level non-nulls, and the other three panels showing which
  groups in each partition are hence identified as non-null. On the
  bottom row is an example of a set of discoveries, with the leftmost
  panel showing the hypothesis-level rejections, and the other three
  panels showing which groups are correspondingly rejected (light-grey
  for correct rejections, black for false rejections). The false
  discovery proportions (FDP) in each partition are 0.2, 0, 0.33, 0.5
  respectively. The true discovery proportions (empirical power) in
  each partition are 0.8, 0.66, 1, 0.5 respectively.}
\label{fig:partitions}
\end{figure}

The various kinds of prior information considered in this paper have been studied in
earlier works and repeatedly motivated in applied settings, 
and our focus is accordingly on the conceptual and
mathematical aspects of multiple decision-making with prior knowledge.
Before beginning our formal presentation, we consider a
simple example, illustrated in Figure~\ref{fig:partitions}, in order
to provide intuition.

Consider a set of sixteen hypotheses arranged in a $4\times4$ grid, as
displayed in the first panel of Figure~\ref{fig:partitions}.  One may
imagine that one coordinate refers to spatial locations, and the other
to temporal locations, so that each square represents an elementary
null hypothesis $H_{s,t}$, stating that there is nothing of interest
occurring at spatial location $s$ at time $t$. As displayed in the
leftmost panel of the first row, the non-nulls may be expected to have
some spatio-temporal structure. In order to exploit this structure,
the scientist may choose to group the hypotheses a priori in three
ways: by spatial location, by temporal location, and by
spatio-temporal blocks, as displayed by the other three panels in the
first row. Each such group can be associated with a group null hypothesis,
which states that there is nothing of interest occurring within that
group.

As displayed in the first row, the group-level non-nulls are simply
implied by the elementary non-nulls.  The second row of the figure
displays the results of a hypothetical procedure that makes some
elementary rejections (first panel), and hence some corresponding
group level rejections (other three panels). The scientist may wish to
not report too many false elementary rejections, but also not report
discoveries at spurious locations, times or space-time blocks. One way
of enabling this wish is to enforce group FDR constraints, in addition
to overall FDR control. This would correspond to controlling the
``spatial'' FDR, ``temporal'' FDR, and ``spatio-temporal'' FDR, in
addition to the overall FDR. Requiring the rejected hypotheses to
satisfy additional constraints may reduce power, but it may also
increase interpretability, and result in higher precision (achieved
FDP). 

For example, \citet{barber2016p} consider an example from neuroscience where 
each null hypothesis $H_{v,t}$ states that at $t$ seconds after the presentation of the stimulus, 
a chosen feature of 
the stimulus is unrelated to (a specific measure of) the brain activity at voxel $v$ in the brain. 
One may spatially group these hypotheses according to pre-defined regions of interest (ROIs) such as
the  visual cortices V1 to V5 and the left/right temporal cortices. This spatial partition allows us to capture the idea that either
an ROI is unrelated to the stimulus, or many of its voxels $v$ will be related. Similarly, for a fixed voxel, one may temporally group the hypotheses, to capture the idea that either a voxel $v$ will remain unrelated to the stimulus at various delays, or it will be related at several consecutive delays (usually $t=4,6,8,10$ seconds after stimulus onset).

The earlier \pf~algorithm \cite{barber2016p} simultaneously
guarantees FDR control for multiple arbitrary and possibly
non-hierarchical partitions, while ensuring that the elementary and
group rejections are ``internally consistent.'' It is a
multi-dimensional step-up procedure, one which reduces to the BH
step-up procedure or Simes' GN test in special cases.  
The reader may
refer to the original paper for numerical simulations and more details on the neuroscience example.


\subsection*{Our contributions}




Consider a collection $\{H_1, \ldots, H_n \}$ of $n$ unordered
hypotheses, along with associated $p$-values $\{P_1,\dots,P_n\}$.  It is
convenient to introduce the shortand notation $[n] \defn
\{1,\dots,n\}$. Further, consider $M$ arbitrary unordered partitions
of these hypotheses into groups, where the $m$-th partition
(``layer'') contains an unordered set of $G^{(m)}$ groups:
\begin{align*}
\small A^{(m)}_1,\dots,A^{(m)}_{G^{(m)}}\subseteq [n] \text{ for
  $m=1,\dots,M$.}
\end{align*}
It may help the reader to imagine the first partition as being the
elementary or finest partition, meaning that it contains $n$ groups of
one hypothesis each, but it is important to note that this partition
is entirely optional, and can be dropped if there is no desire of
controlling the overall FDR. If $\nulls \subseteq n$ are the true
nulls, then we call a group $g$ null if $A^{(m)}_g \subseteq
\nulls$. Let the set of null groups in partition $m$ be denoted by
$\nullsm$.

Although we continue to use the name \pf~for the algorithm that
we discuss in this paper, the algorithm goes significantly beyond the
original algorithm; in particular, in our general setting of $M$
arbitrary partitions there are seven ways in which the new procedure
to be developed here goes beyond the original framework.

\begin{enumerate}

\item \textbf{Overlapping groups.} We allow the groups in any
  partition to overlap. An elementary hypothesis need not be part of
  just a single group \mbox{$g \in [G^{(m)}]$,} and we let $g^{(m)}(i)$ denote
  the set of groups in the $m$-th partition to which $P_i$ belongs---viz.
\begin{align*}
  g^{(m)}(i) = \{g \in [G^{(m)}] : P_i \in A^{(m)}_g \}.
\end{align*}
For example, in the neuroscience example introduced earlier, 
if the scientist is
unsure about the accuracy of the ROI borders, they may place boundary
hypotheses into two or more ROIs to reflect this uncertainty.

\item \textbf{Incomplete partitions.} We allow partitions to be
  incomplete---let the $m$-th partition's leftover set
  $\L^{(m)} \subset [n]$ represent all elementary hypotheses that do
  not belong to any group in the $m$-th partition:
\begin{align*}
\small \L^{(m)} = [n] \backslash \bigcup_g A^{(m)}_g 
\end{align*}
This gives additional flexibility to the user who may not want to
assign some hypotheses to any groups. Note that $\L^{(m)}$ is not just
another group; this set is not counted when calculating the
group-level FDR in layer $m$, meaning that elementary discoveries
within $\L^{(m)}$ do not alter the group FDR at layer $m$. Hence,
hypotheses in this leftover set have no internal consistency
constraints imposed by layer $m$. For instance, in the neuroscience example,
if it is determined (for example due to brain damage or surgery) 
that some voxels may not naturally fit into any ROI, 
then they can be left out of that partition.

\item \textbf{Internal consistency (IC)}. In order to maintain
  interpretability when dealing with overlapping groups, it is
  convenient to introduce two natural notions of internal consistency
  of the group rejections and elementary rejections:
\begin{itemize}
\item \textbf{Weak IC}. We reject $H_i$ if and only if in
every partition, either there is at least one rejected group containing $i$, or $i\in L^{(m)}$.
\item \textbf{Strong IC}. We reject $H_i$ if and only if in
every partition, either every group that contains $i$ is rejected, or $i\in L^{(m)}$.
\end{itemize}
These definitions\footnote{We remark these are not the only two
notions of internal consistency that can fit into our framework: any
\emph{monotone} notion of IC can be handled, where \emph{monotone}
means that decreasing the $p$-values can only possibly increase the
number of rejections at all layers.} of IC are extensions to the multilayer setting of
the notions of \emph{coherence} and \emph{consonance} as defined by
\citet{gabriel1969simultaneous}, and explored in the FWER literature
by Sonnemann and
Finner~\cite{sonnemann1982allgemeine,sonnemann2008general,sonnemann1988vollstandigkeitssatze},
and \citet{romano2011consonance}. In the aforementioned
neuroscience example, weak internal consistency may be more
appropriate.

\item \textbf{Weights.}  The $m$-th partition can be associated with two
  sets of positive weights, one pair for each group $g$ in that
  partition:
\begin{align*}
\text{Penalties } \{u^{(m)}_g\} \text{ and priors } \{w^{(m)}_g\},
\text{ such that } \sum_{g=1}^{G^{(m)}} u^{(m)}_g w^{(m)}_g = G^{(m)}.
\end{align*}
This generalizes work on doubly-weighted procedures by
\citet{blanchard2008two}, who considered a single partition.  Their
work in turn generalizes earlier work using prior weights
\cite{genovese2006false} and penalty weights \citet{BH97} separately.
Large prior weights indicate beliefs that the hypotheses are more
likely to be non-null, and large penalties reflect which hypotheses
are more scientifically important. For example, in the neuroscience example,
weights can also be used to take differing ROI sizes into account, 
or prior knowledge of when and where effects are expected to be found.
\item \textbf{Reshaping.}  Reshaping functions $\beta$ are used to
  guard against possible dependence among the $p$-values by
  \emph{undercounting} the size (or weight) of rejected sets.
  Reshaping makes it possible to handle arbitrary dependence; on the
  other hand, this favorable robustness property is accompanied by a
  loss of power.  Following Blanchard and Roquain~\citep{blanchard2008two}, for any probability measure $\nu$ on
  $[0,\infty)$, we define the reshaping function
\begin{align}
\label{eqn:reshaping}
\beta(k) \defn \int_{0}^k x~ \mathsf{d}\nu(x) ~\leq~ k.
\end{align}
If the $p$-values within or across layers are arbitarily dependent, we
may use reshaping functions $\beta^{(m)}$ to reshape thresholds in
layer $m$.  In the special case that Simes $p$-values are used to form
the group-level $p$-values $P^{(m)}_g$, we may use reshaping functions
$\tbeta^{(m)}_g$ to protect the Simes procedure from arbitrary
dependence within the group.  The original procedure of \citet{BY01}
corresponds to choosing the reshaping function $\beta_{BY}(k) =
\frac{k}{\sum_{i=1}^n \frac1i}$.  Many other examples and their
connections to other formulations of multiple testing methods can be
found in the
literature~\citep{blanchard2008two,sarkar2008methods,sarkar2008two}.
In contrast to the discrete distributions which have been the focus of
past work, in the current paper it is necessary to consider continuous
measures since the penalty weight of rejected hypotheses, unlike their
count, can be fractional.

%
\item \textbf{Adaptivity.} For any partition whose group $p$-values are known to
  be independent (i.e., independence \emph{between} groups, but not necessarily
  \emph{within} each group), we can incorporate ``null-proportion
  adaptivity'' for that partition \cite{hochberg1990more,benjamini2000adaptive}.  For partition $m$, we fix a user-defined constant $\lambda^{(m)} \in (0,1)$,
  and define a weighted null-proportion estimator:

  \begin{small}
\begin{align}
    \label{eqn:pihatm}
\pih^{(m)} := \frac{|u^{(m)} \cdot w^{(m)}|_\infty + \sum_{g}
  u^{(m)}_g w^{(m)}_g \One{P^{(m)}_g >
    \lambda^{(m)}}}{G^{(m)} (1-\lambda^{(m)})} \;.  
\end{align}
\end{small}
The use of null-proportion adaptivity in any \emph{one layer} may improve
the power in \emph{all layers}, since more groups being discovered in
one layer leads to more elementary discoveries, and
hence more discovered groups  in other layers. 
For a single group with no weights, our approach reduces to the original suggestion 
of Storey et al.~\cite{Storey02,Storey04}.
  \item \textbf{Arbitrary group $p$-values.} Our new $\pf$ algorithm 
is no longer tied to the use of Simes $p$-values at the group layers, unlike the original algorithm. 
In other words, each group-level $p$-value at each layer can
be formed by combining the elementary $p$-values within that group~\cite{vovk2012combining,heard2018choosing}.
When the $p$-values are independent,
some options include Fisher's $-2\sum_i \ln P_i$ and Rosenthal's
$\sum_i \Phi^{-1}(P_i)$, where $\Phi$ is the Gaussian CDF (originally proposed by \citet{stouffer1949american}). When
there are very few non-nulls, the Bonferroni correction is known to be
more powerful, and it also works under arbitrary dependence, 
as does R\"uschendorf's proposal of $2 \sum_i P_i / n$, and R\"uger's proposal of $P_{(k)} \cdot n/k$ for a fixed $k$.
Alternately the group $p$-values can be constructed directly from raw data.
Accordingly, we can appropriately use
adaptivity  or reshaping, as needed, depending on the induced dependence. 
%
\end{enumerate}

Suppose a procedure rejects a subset $\Sh\subseteq[n]$ of hypotheses and a
subset $\Shm \subseteq[G^{(m)}]$ of groups in partition $m$.
Then, we may define the penalty-weighted FDR as
\begin{align*}
\fdr^{(m)}_u = \EE{\dotfrac{\sum_{g \in \nullsm} u^{(m)}_g \One{g \in \Shm}}{\sum_{g \in [G^{(m)}]} u^{(m)}_g \One{g \in \Shm}}} \;.
\end{align*}
The rejections made by \pf~will be internally consistent, and satisfy
\begin{align*}
\fdr^{(m)}_u \leq \alpha^{(m)} \text{ simultaneously for all } m = 1,\dots,M.
\end{align*}
In order to handle the possibility of a ratio ``$\frac{0}{0}$'' in this
definition, and in our later work, we adopt the ``dotfraction'' notation 
\begin{align}
\label{eqn:define_dotfrac}
\dotfrac{a}{b} & \defn
\begin{cases}
0,&\text{ if } a=0,\\ 
\frac{a}{b},&\text{ if }a \neq 0,b\neq 0,\\ 
\text{undefined } &\text{ if }a\neq 0, b=0.
\end{cases}
\end{align}
Dotfractions behave like fractions whenever the denominator is nonzero. 
The use of dotfractions simplifies the presentation:
note that $\dotfrac{a}{b} \neq \frac{a}{\max(b,1)}$ since $b$ may 
be fractional due to the use of penalty weights.
We formally derive properties of dotfractions in the supplement (\appref{dotfrac}).

When there is only one partition, and the weights equal one, the
quantity $\fdr^{(m)}_u$ reduces to the usual FDR
defined by \citet{BH95}, and $\pf$~reduces to the BH
procedure\footnote{For a review of its history that involves Eklund and Seeger in the 1960s, and Simes, Hommel, Soric, Benjamini and Hochberg in the 1980s and 1990s, see \cite{seeger1968note,benjamini2000adaptive}.}. Many other
procedures are recovered as special cases of $\pf$, as
detailed after \thmref{pf+}.

\section{\texorpdfstring{$\fdr$}{Lg} control and internal consistency for multiple partitions}\label{sec:pf+}

Even in the case of just two partitions---one partition of groups and
the elementary partition of individual hypotheses---it is non-trivial
to provide a guarantee of internal consistency while controlling both
group-level and individual-level FDR.  For example, a sequential
procedure, of first rejecting groups at a target FDR level $\alpha_1$
and then rejecting individual hypotheses within rejected groups at
level $\alpha_2$, may not control the elementary FDR (due to not
accounting for selection bias), and may not be internally consistent
(because there might be a group rejected in the first round, 
with none of the elementary hypotheses in this group rejected in the second round). 
Further, such a method is not
easily generalized to non-hierarchical partitions.  Similarly, a
parallel procedure that independently runs FDR procedures on the
groups and on the individuals, may also fail to be internally
consistent.  Naively intersecting these rejections---that is,
rejecting those hypotheses whose groups are rejected at every
layer---may also fail to control the FDR (see \citet{barber2016p} for
explicit examples).

The \pf~algorithm is a multivariate extension of classical step-up procedures
 that is roughly based on the following sequence of steps:
\begin{itemize}
\item Select all hypotheses in each layer whose $p$-values are smaller than some 
initial layer-specific threshold.

\item Reject an elementary hypothesis if it is contained in a selected group in every layer. 

\item In each layer, reject a group hypothesis if it contains a rejected elementary hypothesis. 
Then, estimate the group-FDP in each layer.
\item Lower the initial thresholds at each layer, and repeat the steps above, until the group-FDP is below the desired level 
for all partitions.
\end{itemize}

 Next, we discuss the necessary dependence
 assumptions and then derive the \pf~algorithm that implements the above
 scheme.

\subsection{Marginal and joint distributional assumptions on $p$-values}

We assume that the marginal distribution of each null
$p$-value is stochastically
larger than the uniform distribution, referred to as \emph{super-uniform} for brevity.  
Formally, for any index
$i \in \nulls$, we assume that
\begin{align}
  \label{EqnSuperUniform}
\PP{P_i \leq t} \leq t \quad\mbox{for all $t \in [0,1]$.}
\end{align}
Of course, uniformly-distributed $p$-values trivially satisfy this condition.
We use the phrase \emph{under uniformity} to
  describe the situation in which the null $p$-values are marginally
exactly  uniform. If this phrase is not employed, it is understood
  that the null $p$-values are marginally super-uniform.

Regarding assumptions on the joint distribution of $p$-values, three
possible kinds of dependence will be considered in this paper:
independence, positive dependence or arbitrary dependence.  
In the independent setting, null $p$-values are assumed to be mutually
independent, and independent of non-nulls. In the arbitrary dependence setting, 
no joint dependence assumptions are made on the $p$-values.
The last case is that of positive dependence, as formalized by the
\emph{Positive Regression Dependence on a Subset} (PRDS)
condition~\citep{lehmann1966some,sarkar1969some,BY01}.  In order to
understand its definition, it is helpful to introduce some basic
notation.  For a pair of vectors $x, y \in [0,1]^n$, we use the
notation $x \preceq y$ to mean that $x \leq y$ in the orthant
ordering, i.e., $x_i \leq y_i$ for all $i \in \{1, \dots, n\}$.  A set
$\Dset \subseteq [0,1]^n$ is said to be \emph{nondecreasing} if $x \in
\Dset$ implies $y \in \Dset$ for all $y \succeq x$. We say that a
function $f: [0,1]^n \mapsto \Rplus$ is \emph{nonincreasing}, if $x
\preceq y$ implies $f(x) \geq f(y)$.
\begin{definition}[positive dependence, PRDS]
\label{ass:PRDS}
We say that the vector $P$ satisfies PRDS if for any null index $i \in
\nulls$ and nondecreasing set $ \Dset \subseteq [0,1]^n$, the function
$t~\mapsto~\PPst{P\in \Dset}{P_i \leq t}$ is nondecreasing over
$t\in(0,1]$.
\end{definition}
The original positive regression dependence assumption was introduced
by~\citet{lehmann1966some} in the bivariate setting and
by~\citet{sarkar1969some} in the multivariate setting, and extended to
the PRDS assumption first made by~\citet{BY01}.  These previous papers
used the equality $P_i = t$ instead of the inequality $P_i \leq t$ in
the definitions, but one can prove that both conditions are
essentially equivalent.

The PRDS condition holds trivially if the $p$-values are independent,
but also allows for some amount of positive dependence.  For
intuition, suppose that \mbox{$Z = (Z_1,\ldots,Z_n)$} is a
multivariate Gaussian vector with covariance matrix $\Sigma$; the null
components correspond to Gaussian variables with zero mean.  Letting
$\Phi$ be the CDF of a standard Gaussian, the vector of $p$-values
\mbox{$P=(\Phi(Z_1),\dots,\Phi(Z_n))$} is PRDS on $P_i$ for every
index $i$ if and only if all entries of the covariance matrix $\Sigma$
are non-negative.  See~\citet{BY01} for additional examples of this
type.

\subsection{Specifying the \texorpdfstring{$\pf$}{Lg} algorithm}

In order to run the $\pf$ algorithm, we need to search for rejection
thresholds for each layer. These thresholds will be parametrized by
\emph{weighted} discovery counts $k^{(m)}\in[0,G^{(m)}]$ for each
layer $m=1,\dots,M$. The reader is cautioned that each $k^{(m)}$ need
not be an integer but instead should be viewed as a real number
corresponding to the total rejected penalty weight.  If the weights
$u^{(m)}_g$ are all set equal to one, then $k^{(m)}$ corresponds to
the number of groups in layer $m$ that are rejected.  Given some
prototypical vector $\vec{k} \defn (k^{(1)},\dots,k^{(M)})$, we first
perform an initial screening on each layer separately:
  \begin{align}
    \label{eqn:Shm}
\Shm_{\textnormal{init}}(\vec{k}) = \left\{g\in[G^{(m)}]: P^{(m)}_g
\leq \min \Big \{ \tfrac{ w^{(m)}_{g} \alpha^{(m)}
  \beta^{(m)}(k^{(m)})}{\pih^{(m)} G^{(m)}}, \lambda^{(m)} \Big \}
\right\}.
\end{align}
If the groups in partition $m$ are independent, we replace
$\beta^{(m)}(k^{(m)})$ by just $k^{(m)}$, and set $\pih^{(m)}$
using~\eqref{eqn:pihatm}; on the other hand, if they are arbitrarily
dependent, we set $\pih^{(m)}=1$ and $\lambda^{(m)}=1$.  This
convention allows the same expressions to be used in all settings.

For weak internal consistency, we define the elementary rejections as
\begin{subequations}
  \begin{align}
    \label{eqn:hatS-defn}
\Sh(\vec{k}) & = \Sh_{\textnormal{weak}}(\vec{k})= \bigcap_{m=1}^M
\Biggr( \Big[\bigcup_{g\in \Shm_{\textnormal{init}}(\vec{k})}
  A^{(m)}_g \Big] \cup L^{(m)} \Biggr) \nonumber \\
&= \{ P_i : \text{$\forall m$, either } P_i \in L^{(m)}, \text{ or
  $\exists \ g \in g^{(m)}(i)$, } A^{(m)}_g \in
\Shm_{\textnormal{init}}(\vec{k}) \}.
\end{align}
Alternately, for strong internal consistency, we may instead define
\begin{align}\label{eqn:hatS-defn-strong}
\Sh(\vec{k}) &= \Sh_{\textnormal{strong}}(\vec{k})= \bigcap_{m=1}^M
\left([n]\backslash \bigcup_{g\in[G^{(m)}]\backslash
  \Shm_{\textnormal{init}}(\vec{k})} A^{(m)}_g\right) \nonumber \\
& = \{ P_i : \text{$\forall m$, either } P_i \in L^{(m)}, \text{ or
  $\forall \ g \in g^{(m)}(i)$, } A^{(m)}_g \in
\Shm_{\textnormal{init}}(\vec{k}) \}.
\end{align}
\end{subequations}

Finally, using either $\Sh(\vec{k})=\Sh_{\textnormal{weak}}(\vec{k})$
or $\Sh(\vec{k})=\Sh_{\textnormal{strong}}(\vec{k})$, we redefine the
set of groups in layer $m$ which are rejected as:
\begin{align}
  \label{eqn:hatSm-defn}
\Shm(\vec{k}) = \left\{ g \in [G^{(m)}]: A^{(m)}_g\cap \Sh(\vec{k})
\neq \emptyset \text{ and } g \in \Shm_{\textnormal{init}}(\vec{k})
\right \}.
\end{align}
Examining these definitions, it may be verified that (weak or strong)
internal consistency is satisfied by the rejections $\Sh(\vec{k}),
\Sh_m(\vec{k})$.

Of course, these definitions depend on the initial choice of the
vector $\vec{k}$. Since we would like to make a large number of
discoveries, we would like to use a $\vec{k}$ that is as large as
possible (coordinatewise), while at the same time controlling the
layer-specific FDRs, which are the expectations of
\begin{align*}
  \small \fdpm_u(\vec{k}) \defn \dotfrac{\sum_{g \in \nullsm}
    u^{(m)}_g \One{g \in \Shm(\vec{k})} }{\sum_{g \in [G^{(m)}]}
    u^{(m)}_g \One{g \in \Shm(\vec{k})}} \;.
\end{align*}

\vspace{0.2in}
\noindent Now, define the data-dependent set of feasible vectors
$\vec{k}$ as
\begin{align}
  \label{eqn:feasible-thresholds}
\Kset = \left \{ \vec{k} \in [0,G_1] \times \dots \times [0,G_M] :
\sum_{g \in \Shm(\vec{k})} u^{(m)}_g \geq k^{(m)} \textnormal{ for all
} m \right \},
\end{align}
where we suppress the implicit dependence of $\Kset$ on input
parameters such as $\alpha^{(m)}, \lambda^{(m)},
\{w^{(m)}_g\},\{u^{(m)}_g\}$.  In particular, if the penalty weights
are all equal to one, then the consistency condition defining the
``feasible'' $\vec{k}$'s is equivalent to requiring that
$|\Shm(\vec{k})| \geq k^{(m)}$ for all $m=1,\dots,M$; i.e., the
numbers of rejections in each layer at the vector $\vec{k}$ are
elementwise $\geq \vec{k}$.  This condition can be viewed as a
generalization, to the multi-partition setting, of the
``self-consistency'' condition described by \citet{blanchard2008two}.

It is also worth noting that the $\pf$ algorithm in
\citet{barber2016p} was derived in terms of thresholds $\vec{t}$
instead of number of rejections $\vec{k}$, and there the corresponding
feasiblity condition was that $\fdphat_m(\vec{t}) \leq \alpha^{(m)}$,
where $\fdphat_m(\vec{t})$ is an empirical-Bayes type estimate of the
FDP. Indeed, if we avoid $\pih^{(m)},\beta^{(m)},w^{(m)},u^{(m)}$ for
simplicity, then associating $\th^{(m)}$ to $\alpha^{(m)} \kh^{(m)} /
G^{(m)}$ and comparing our derivation to that of \citet{barber2016p},
we can see that the ``self-consistency'' viewpoint and the
``empirical-Bayes'' viewpoint are equivalent and lead to the same
algorithm. However, when dealing with reshaping under arbitrary
dependence, the proofs are simpler in terms of $\vec{k}$ than in terms
of $\vec{t}$, explaining our switch in notation.

As with the BH and BY procedures, we then choose the largest feasible thresholds $k^{(m)}$, given by:
\begin{align}\label{eqn:max-threshold-m}
\kh^{(m)} = \max\left\{k^{(m)}  :  \exists k^{(1)},\dots,k^{(m-1)},k^{(m+1)},\dots,k^{(M)}\textnormal{ s.t. } \vec{k} \in\Kset \right\} \;. 
\end{align}
This choice defines our algorithm: the $\pf$ algorithm rejects the hypotheses $\Sh(\kh^{(1)},\dots,\kh^{(M)})$, as defined in~\eqnref{hatS-defn} or~\eqnref{hatS-defn-strong}, 
with rejections at layer $m$ given by $\Shm(\kh^{(1)},\dots,\kh^{(M)})$ as defined in~\eqnref{hatSm-defn}.
Next, we present the theoretical guarantees associated with $\pf$.

\subsection{Theoretical guarantees}

The following proposition states that the set of feasible vectors $\Kset$
actually has a well-defined ``maximum'' corner.
\begin{proposition}
  \label{prop:max}
Let the set of feasible vectors $\Kset$ be defined as in
equation~\eqref{eqn:feasible-thresholds}, and let the partition-specific
maximum feasible vector $\kh^{(m)}$ be defined as in
equation~\eqref{eqn:max-threshold-m}. Then we have
\begin{align}
\label{eqn:feasible-thresholds2}
  (\kh^{(1)},\dots,\kh^{(M)})\in\Kset \;.
\end{align}
\end{proposition}
\noindent
The proof is provided in \secref{proof_max}; it is a generalization of
the corresponding result for the original $\pf$
algorithm~\cite{barber2016p}.

The vector $(\kh^{(1)},\dots,\kh^{(M)})$ is not just feasible from the
perspective of self-consistency as captured by $\Kset$, but it is also
feasible from the perspective of FDR control.  Specifically, the next
theorem establishes that---assuming for now that we can find
$(\kh^{(1)},\dots,\kh^{(M)})$---selecting the set
$\Sh(\kh^{(1)},\dots,\kh^{(M)})$ guarantees simultaneous control of
$\fdr_u^{(m)}$ for all $M$ partitions.  In this theorem, the notation
$\simes_w(P_{A^{(m)}_g})$ refers to the weighted Simes' $p$-value (see
\appref{simes} in the supplement for details).
\begin{thm}
\label{thm:pf+}
Any procedure that computes the vector $(\kh^{(1)},\dots,\kh^{(M)})$
according to definition~\eqref{eqn:max-threshold-m} satisfies the
following properties, for all partitions $m = 1,\dots,M$
simultaneously:
\begin{enumerate}
\item[(a)] If the base $p$-values are independent, and all group
  $p$-values are given by $P^{(m)}_g = \simes_w(P_{A^{(m)}_g})$, then
  employing adaptivity by defining $\pih^{(m)}$ as
  in~\eqref{eqn:pihatm} guarantees that
  $\fdr_u^{(m)}\leq\alpha^{(m)}$.
\item[(b)] If base $p$-values are positively dependent (PRDS) and group
  $p$-values are given by $P^{(m)}_g = \simes_w(P_{A^{(m)}_g})$, then
  without adaptivity or reshaping, we have that $\fdr_u^{(m)}\leq
  \alpha^{(m)} \frac{\sum_{g \in \nullsm} u^{(m)}_g
    w^{(m)}_g}{G^{(m)}} \leq \alpha^{(m)}$.
 \item[(c)] When all $p$-values are arbitrarily dependent, and are
   constructed arbitrarily (under the assumption that $P^{(m)}_g$ is
   super-uniform for any null group $g\in\nullsm$, meaning it is a
   valid $p$-value), then using reshaping as in~\eqref{eqn:reshaping}
   guarantees that $\fdr_u^{(m)}\leq \alpha^{(m)}\frac{\sum_{g \in
       \nullsm} u^{(m)}_g w^{(m)}_g}{G^{(m)}} \leq \alpha^{(m)}$.
\item[(d)] In the setting of part (c), if additionally the groups at
  layer $m$ are independent (that is, $P_{A^{(m)}_g}$ is independent
  from $P_{-A^{(m)}_g}$, for each $g\in[G^{(m)}]$), then using
  reshaping as in~\eqref{eqn:reshaping} and adaptivity for layer $m$
  as in~\eqref{eqn:pihatm}, guarantees that
  $\fdr_u^{(m)}\leq\alpha^{(m)}$.
\end{enumerate}
\end{thm}

The proof, given in \secref{proof_main}, uses three interpretable
lemmas that we first discuss in \secref{lemmas}.  It also introduces
several new ideas to handle overlapping groups with dependent
$p$-values.  To remark on the difference between parts (c) and (d), what
these two results guarantee is that if we use adaptivity for some set
$\mathcal{M}_{\textnormal{adapt}}\subset[M]$ of layers, and do not use
adaptivity (i.e.~set $\pih^{(m)}=1$) for the remaining layers, then
FDR control is maintained across {\em all} layers as long as, for each
$m\in\mathcal{M}_{\textnormal{adapt}}$, the layer-specific
independence statement holds---$P_{A^{(m)}_g}$ is independent from
$P_{-A^{(m)}_g}$, for each $g\in[G^{(m)}]$.  If this condition fails
for some $m\in\mathcal{M}_{\textnormal{adapt}}$, the FDR control in
other layers will in fact not be affected.  One application of
statement (d) is when the base $p$-values are independent, there are no
overlapping groups, and group $p$-values are formed using a Fisher,
Rosenthal, or other combinations of the base $p$-values. Recently,
\citet{katsevich17mkf} proved that in case (d), the FDR is controlled
even without using reshaping, albeit at a constant factor larger than
the target level.

In practice, if we have accurate side information about group
structures that the rejected hypotheses likely respect, then we may
significantly improve our \textit{precision}, achieving a lower FDR
than the theoretical bound, without affecting our power much. However,
inaccurate side information may significantly lower our power, since
each $p$-value would have additional misguided constraints to meet.
These issues were explored in simulations by \citet{barber2016p}. 


\paragraph{Special cases}  The setting with a single partition ($M=1$)
recovers a wide variety of known algorithms. Considering only the
finest partition with $n$ groups containing one hypothesis each, the
\pf~ algorithm and associated \thmref{pf+} together recover known
results about (a) the BH procedure of \citet{BH95} when weights, reshaping
and adaptivity are not used, (b) the BY procedure of \citet{BY01} when
reshaping is used, (c) the prior-weighted BH procedure of
\citet{genovese2006false} when only prior weights are used, (d) the
penalty-weighted BH procedure of \citet{BH97} when only penalty
weights are used, (e) the doubly-weighted BH procedure of
\citet{blanchard2008two} when both sets of weights and reshaping are
used, and (f) the Storey-BH procedure of Storey et
al.~\cite{Storey02,Storey04} when only adaptivity is used.

When we instantiate $\pf$ with the coarsest partitions with a single
group containing all $n$ hypotheses, we recover exactly (g) the $\simes$
test \cite{simes1986improved} without weights, and (h) a variant by
\citet{HL94} if prior weights are used.  We recover the results of
(i) the p-filter by \citet{barber2016p} under positive dependence, when we do not use
weights, adaptivty, reshaping, overlapping groups, leftover sets, and
restrict ourselves to Simes' $p$-values.  We also recover a host of new
procedures: for example, while the past literature has not yet shown
how to use either prior or penalty weights together with adaptivity,
\pf~reduces to (j) a doubly-weighted adaptive procedure for the finest
partition under independence. Also, while the aforementioned
procedures were each proved under one form of dependence or the other,
we recover results for all three forms of dependence at one go, with a
single unified proof technique.


\subsection{An efficient implementation}

Although one can employ a brute-force grid search to find
$(\kh^{(1)},\dots,\kh^{(M)})$, the \texttt{p-filter} algorithm
presented in Algorithm~\ref{alg:multi-layer_fdr} is able to find this
vector efficiently using a coordinate-descent style procedure, and is
a strict generalization of the algorithm by the same name in
\citet{barber2016p}.

\begin{algorithm}[h]
\caption{The $\pf$ for multi-layer FDR control}
\label{alg:multi-layer_fdr}
\begin{algorithmic}
\STATE \textbf{Input:} $M$ possibly incomplete partitions of possibly
overlapping groups of indices $[n]$;\\\quad\quad\quad A vector of base
$p$-values $P\in[0,1]^n$;\\ \quad\quad\quad Group $p$-values $P^{(m)}_g$
for each group $g=1,\dots,G^{(m)}$ in layers
$m=1,\dots,M$;\\ \quad\quad\quad $M$ target FDR levels
$\{\alpha^{(m)}\}$; \\ \quad\quad\quad $M$ sets of prior weights
and/or penalty weights $\{w^{(m)}_g, u^{(m)}_g\}$;\\ \quad\quad\quad
$M$ thresholds for adaptive null proportion estimation
$\{\lambda^{(m)}\}$;\\ \quad\quad\quad $M$ reshaping functions
$\{\beta^{(m)}\}$, if desired.  \STATE
\textbf{Initialize:} Set $k^{(m)}=G^{(m)}$, and $\pih^{(m)}$ as in
definition~\eqref{eqn:pihatm}.  \REPEAT \FOR{$m=1,\dots,M$} \STATE
Update the $m$th vector: defining $\Shm(\vec{k})$ as in
equation~\eqref{eqn:hatSm-defn} (using weak or strong internal
consistency, as desired), let
\begin{align*}
k^{(m)} \leftarrow \max\left\{k'^{(m)} \in [0,G^{(m)}]: \sum\limits_{g
  \in \Shm(k^{(1)},\dots,k^{(m-1)},k'^{(m)},k^{(m+1)},\dots,k^{(M)})}
u^{(m)}_g \geq k'^{(m)} \right\}
\end{align*}
\ENDFOR \UNTIL{the vectors $k^{(1)},\dots,k^{(M)}$ are all unchanged
  for one full cycle.}  \STATE \textbf{Output:} Vector
$\kh=(k^{(1)},\dots,k^{(m)})$, rejected hypotheses $\Sh(\kh)$, and
rejected groups $\Shm(\kh)$ in each partition.
\end{algorithmic}
\end{algorithm}

The following proposition provides a correctness guarantee for
Algorithm~\ref{alg:multi-layer_fdr}:
\begin{proposition}
  \label{prop:alg}
The output of Algorithm~\ref{alg:multi-layer_fdr} is the maximum
feasible corner $(\kh^{(1)},\dots,\kh^{(m)})$ defined in
equations~\eqref{eqn:max-threshold-m}
and~\eqref{eqn:feasible-thresholds2}.
\end{proposition}
\noindent This result was proved by \citet{barber2016p} in the setting
of the original $\pf$ algorithm, where the $k^{(m)}$'s take only
integer values; here, the algorithm is slightly more subtle, with
real-valued $k^{(m)}$'s due to the presence of penalty weights
$u^{(m)}_g$.  The proof of the proposition for this more general
setting is given in \secref{proof_alg}.

\section{Three lemmas}
\label{sec:lemmas}

In this section, we present three lemmas that lie at the heart of the
succinct proofs of the theorems in this paper.  Our motivation for
presenting these lemmas here is that they are interpretable, and provide
valuable intuition for the proofs that follow.


\subsection{A super-uniformity lemma for FDR control}

In order to develop some intuition for the lemma that follows, we note
that our super-uniformity assumption~\eqref{EqnSuperUniform} on the
null $p$-values can be reformulated as:
\begin{align}
\label{eqn:PRDS-fixed}
\text{For any $i \in \nulls$, ~}~ \EE{\dotfrac{{\One{P_i\leq t}}}{t}}
& \leq 1 \text{ for any fixed $t \in [0,1]$.}
\end{align}
Of course, if $P_i$ is uniform then the above inequality holds with
equality.

The following lemma guarantees that property~\eqnref{PRDS-fixed}
continues to hold for certain random thresholds $f(P)$. Recall that
the term ``nonincreasing'' is interpreted coordinatewise, with respect
to the orthant ordering.

\begin{lemma}[Super-uniformity lemma]
\label{lem:power}
Let $i \in \nulls$ be a null hypothesis with $p$-value $P_i$, and let
$P^{-i}$ denote the other $n-1$ $p$-values.
  \begin{enumerate}
  \item[(a)] For any nonincreasing function
    $f:[0,1]^n\rightarrow[0,\infty)$, if $P_i$ is independent of
    $P^{-i}$, then we have
\begin{align*}
 \EEst{\dotfrac{\One{P_i\leq f(P)}}{f(P)} }{ P^{-i}} \leq 1.
\end{align*}
Furthermore, if we additionally assume that $f$ has range $[0,1]$ and
satisfies the LOOP condition (supplement, \appref{LOOP}), and that $P_i$ is
uniformly distributed, then the inequality is replaced with equality:
\begin{align*}
 \EEst{\dotfrac{\One{P_i\leq f(P)}}{f(P)} }{ P^{-i}} = 1.
\end{align*}
   \item[(b)] For any nonincreasing function
     $f:[0,1]^n\rightarrow[0,\infty)$, if $P$ is PRDS with respect to
     $P_i$, then
\begin{align*}
\EE{\dotfrac{\One{P_i\leq f(P)}}{f(P)} } \leq 1.
\end{align*}
\item[(c)] For any constant $c\geq 0$, any function
  $f:[0,1]^n\rightarrow[0,\infty)$, and any reshaping function
  $\beta$, under arbitrary dependence of the $p$-values,
\begin{align*}
\small \EE{\dotfrac{\One{P_i\leq c\cdot \beta(f(P)) }}{c\cdot f(P)} }
\leq 1.
\end{align*}
\item[(d)] For any constant $c\geq 0$, any functions
  $f_1,\dots,f_m:[0,1]^n\rightarrow[0,\infty)$, and any reshaping
  functions $\beta_1,\dots,\beta^{(m)}$, under arbitrary dependence of
  the $p$-values,
\begin{align*}
\small \EE{\dotfrac{\One{P_i\leq c\cdot
      \prod_{\ell=1}^m\beta_\ell(f_\ell(P)) }}{c\cdot \prod_{\ell=1}^m
    f_\ell(P)} } \leq 1 .
\end{align*}
\end{enumerate}
\end{lemma}
The proofs of statement (a) with equality, and of statement (d), are
given in \secref{power-proof}. Statement (a) with inequality is
recovered as a special case of statement (b), which was proved by
\citet{blanchard2008two}, who also proved (c).  The more general
statement (d), with more than one reshaping function present in the
bound, will be required in the proof of the following novel group
super-uniformity \lemref{power3}.


\subsection{A group-level super-uniformity lemma}

In analogy to the super-uniformity \lemref{power}, we present the
following lemma, which contains analogous bounds under the settings of
independent or positively dependent base $p$-values (in which case the
group $p$-value is constructed with a Simes $p$-value), and in the setting
of arbitrarily dependent base $p$-values (in which case the group
$p$-value can be constructed by any method---reshaped Simes, Fisher, or
others---as long as it is a valid $p$-value.)

\begin{lemma}[Group-level super-uniformity lemma]
\label{lem:power3} 
Let $g \in \nullsg$ be a null group, that is, $A_g\subseteq
\nulls$. Let $P_{A_g}$ denote the $p$-values in this group,
$P_{A_g}=(P_j)_{j\in A_g}$, and let $P_{-A_g}$ denote the remaining
$p$-values, $P_{-A_g}=(P_j)_{j\not \in A_g}$.
  \begin{enumerate}
    \item[(a)] If $f:[0,1]^n\rightarrow[0,\infty)$ is a nonincreasing
      function, and the base $p$-values $P_1,\dots,P_n$ are independent,
      then
\begin{align*}
\EEst{\dotfrac{\One{\simes_w(P_{A_g})\leq f(P)}}{f(P)} }{P_{-A_g}} \leq 1.
\end{align*}
    \item[(b)] If $f:[0,1]^n\rightarrow[0,\infty)$ is a nonincreasing
      function, and the base $p$-values $P_1,\dots,P_n$ are positively
      dependent (PRDS), then
\begin{align*}
\EE{\dotfrac{\One{\simes_w(P_{A_g})\leq f(P)}}{f(P)} } \leq 1.
\end{align*}
\item[(c)] If the base $p$-values $P_1,\dots,P_n$ are arbitrarily
  dependent, then for any constant $c>0$, any reshaping function
  $\beta$, and any function $f:[0,1]^n\rightarrow[0,\infty)$, we have
\begin{align*}
\EE{\dotfrac{\One{T(P_{A_g})\leq c \beta(f(P)) }}{cf(P)} } \leq 1,
\end{align*}
where $T:[0,1]^{|A_g|}\rightarrow[0,1]$ is any valid group $p$-value;
i.e., any function with the property that $T(P_{A_g})$ is
super-uniform whenever $g$ is null.
\item[(d)] Let $g_1,\dots,g_k$ be a set of $k$ possibly overlapping
  null groups, meaning $A_{g_1},\dots,A_{g_k} \subseteq \nulls$, and
  $S_1,\dots,S_k$ represent the corresponding Simes' $p$-values. If
  $f:[0,1]^n\rightarrow[0,\infty)$ is a nonincreasing function, and
    the base $p$-values $P_1,\dots,P_n$ are positively dependent (PRDS), then
\begin{align*}
\EE{\dotfrac{\One{\simes(S_1,\dots,S_k) \leq f(P)}}{f(P)} } \leq 1.
\end{align*}
\end{enumerate}
\end{lemma}
\noindent The proof of this lemma relies on \lemref{power}, and can be
found in \secref{power2-proof}. We remark that statement (d) is
different from statement (b) applied to the null group $g =
\bigcup_{i=1}^k g_i$; indeed, in statement (d), the arguments to the
Simes' procedure are themselves Simes' $p$-values, and not the original
base $p$-values.  If desired, statement (d) can be further bootstrapped
to apply to the root of an entire tree of null groups, where each
internal node stores the Simes' $p$-value calculated on its children.

As an aside, one may wonder whether the Simes' $p$-values are themselves
positively dependent (PRDS), given that they satisfy a
super-uniformity lemma much like the PRDS $p$-values. We have neither
been able to prove nor disprove such a claim, and it may be of
independent interest to do so.


\subsection{An inverse binomial lemma for adaptivity with weights}

The following lemma is required for the proof of adaptivity with
weights; more specifically, we use it to bound the expected inverse of
the doubly-weighted null-proportion estimate.

\begin{lemma}[Inverse binomial lemma]
\label{lem:power2}
Given a vector $a \in [0,1]^{d}$, constant $b \in [0,1]$, and
Bernoulli variables $Z_i ~\stackrel{\textnormal{i.i.d.}}{\sim}~
\text{Bernoulli}(b)$, the weighted sum $Z \defn 1 + \sum_{i=1}^{d} a_i
Z_i$ satisfies
\begin{align}
\frac{1}{1 + b \sum_{i=1}^d a_i } \leq \EE{\frac{1}{Z}} & \leq
\frac{1}{b (1 + \sum_{i=1}^d a_i)}.
\end{align}
\end{lemma}
\noindent Since $\EE{Z} = 1 + b \sum_{i=1}^d a_i $, the lower bound on
$\EE{1/Z}$ follows by Jensen's inequality. We include this bound to
provide context for the upper bound on $\EE{1/Z}$, whose proof can be
found in~\secref{power3-proof}. When $a_i=1$ for all $i$ and $b=1$,
the claim follows by a standard property of binomial distributions, as
described in \citet{benjamini2006adaptive}.\\

With these three lemmas in place, we now turn to the proof of the main
theorem in the next section.

\section{Proof of Theorem~\lowercase{\ref{thm:pf+}}}
\label{sec:proof_main}

\noindent 
In order to be able to handle all four cases of the theorem, we define
a function $\gamma^{(m)}$ to be the identity if we are not using
reshaping (theorem statements (a,b)), or $\gamma^{(m)}=\beta^{(m)}$ if
we are using reshaping (theorem statements (c,d)).  We also let
$\pih^{(m)}=1$ and $\lambda^{(m)}=1$ if we are not using adaptivity
(theorem statements (b,c)), or let $\pih^{(m)}$ be defined as in
equation~\eqref{eqn:pihatm} where adaptivity is used (theorem
statements (a,d)).

Fix any partition $m$.  Since $\PP{P_i=0}=0$ for any $i\in\nulls$ by
assumption, we assume that $P_i\neq 0$ for any $i\in\nulls$ without
further mention; this assumption then implies that if $
g\in\Shm(\kh^{(1)},\dots,\kh^{(M)})$ for some null group $
g\in\nullsm$, we must have $\kh^{(m)}>0$.  We can then calculate
\begin{align*}
\fdpm_u(\kh^{(1)},\dots,\kh^{(M)}) &= \dotfrac{\sum_{g \in \nullsm}
  u^{(m)}_g \One{g \in \Shm(\kh^{(1)},\dots,\kh^{(M)})}}{ \sum_{g \in
    [G^{(m)}]} u^{(m)}_g \One{g \in \Shm(\kh^{(1)},\dots,\kh^{(M)})} }
\\[1.5em] & \leq \dotfrac{\sum_{g \in \nullsm} u^{(m)}_g \One{g \in
    \Shm(\kh^{(1)},\dots,\kh^{(M)})}}{ \kh^{(m)} }, \\[1em] & \leq
\dotfrac{\sum_{g \in \nullsm} u^{(m)}_g \One{g \in
    \Shm_{\textnormal{init}}(\kh^{(1)},\dots,\kh^{(M)})}}{ \kh^{(m)}
}, \\[1em] & = \dotfrac{\sum_{g \in \nullsm} u^{(m)}_g \One{P^{(m)}_g
    \leq \min\{w^{(m)}_g \frac{\alpha^{(m)}
      \gamma^{(m)}(\kh^{(m)})}{\pih^{(m)} G^{(m)}}, \lambda^{(m)} \} }
}{ \kh^{(m)} } ,
\end{align*} 
where the first inequality follows by
definition~\eqref{eqn:feasible-thresholds} of the feasible set
$\Kset$, the second follows since
$\Shm(\vec{k})\subseteq\Shm_{\textnormal{init}}(\vec{k})$ for any
$\vec{k}$ by definition, and the last step uses the definition of
$\Shm_{\textnormal{init}}(\vec{k})$ in~\eqref{eqn:Shm} (without
reshaping, for theorem statements (a,b), or with reshaping for theorem
statements (c,d)).

Multiplying the numerator and denominator of each term by
$\frac{\alpha^{(m)} w^{(m)}_g}{G^{(m)}}$, and taking expectations on
both sides, it follows that

\begin{align}
\label{eqn:maintheorem-midproof-new}
 \textstyle \fdr^{(m)}_u \leq \frac{\alpha^{(m)}}{G^{(m)}} \sum_{g \in
   \nullsm} u^{(m)}_g w^{(m)}_g \EE{\dotfrac{ \One{P^{(m)}_g \leq
       \min\{ \frac{w^{(m)}_g \alpha^{(m)}
         \gamma^{(m)}(\kh^{(m)})}{\pih^{(m)} G^{(m)}}, \lambda^{(m)}
       \} }}{ \frac{w^{(m)}_g \alpha^{(m)} \kh^{(m)}}{ G^{(m)}} }}.
\end{align}

With these calculations in place, we now prove the four statements of
the theorem.  Given the suggestive form of the above expression, it is
natural to anticipate the use of the two super-uniformity lemmas.

\noindent \paragraph{Theorem statement (a)} 

Define the function $f^{(m)}_{g}$ that maps the vector $P$ to
$\frac{w^{(m)}_g \alpha^{(m)} \kh^{(m)} }{\pih^{(m)} G^{(m)}}$.  Note
that $f^{(m)}_{g}$ is a nonincreasing function of $P$, since
$\kh^{(m)}$ is a nonincreasing function of $P$ by definition of our
procedure, while $\pih^{(m)}$ is a nondecreasing function of $P$.  We
also define the quantity
\begin{align}
\label{eqn:pihatm_g}
\pih^{(m)}_{-g} ~\defn~ \frac{|u^{(m)} w^{(m)}|_\infty + \sum_{h \neq
    g} u^{(m)}_h w^{(m)}_h \One{ P^{(m)}_h >\lambda^{(m)}}}{G^{(m)}
  (1-\lambda^{(m)})}.
\end{align}
Returning to expression~\eqref{eqn:maintheorem-midproof-new}, we may
then deduce that
\begin{align}
\fdr^{(m)}_u \notag & \leq \frac{\alpha^{(m)}}{G^{(m)}} \sum_{g \in
  \nullsm} u^{(m)}_g w^{(m)}_g \EE{\dotfrac{ \One{P^{(m)}_g \leq
      \min\{ \frac{w^{(m)}_g \alpha^{(m)} \kh^{(m)}}{\pih^{(m)}
        G^{(m)}}, \lambda^{(m)} \} }}{ \pih^{(m)} \frac{w^{(m)}_g
      \alpha^{(m)} \kh^{(m)}}{\pih^{(m)} G^{(m)}} }}\\ \notag&=
\frac{\alpha^{(m)}}{G^{(m)}} \sum_{g \in \nullsm} u^{(m)}_g w^{(m)}_g
\EE{\One{P^{(m)}_g \leq\lambda^{(m)}}\cdot \dotfrac{ \One{P^{(m)}_g
      \leq f^{(m)}_g(P)}}{ \pih^{(m)} f^{(m)}_g(P) }} \\
\label{eqn:maintheorem-midproof-new-a}
& \stackrel{(i)}{\leq} \frac{\alpha^{(m)}}{G^{(m)}} \sum_{g \in
  \nullsm} u^{(m)}_g w^{(m)}_g \EE{\dotfrac{ \One{P^{(m)}_g \leq
      f^{(m)}_g(P)}}{ \pih^{(m)}_{-g} f^{(m)}_g(P) }},
\end{align}
where inequality (i) holds because the event $P^{(m)}_g\leq
\lambda^{(m)}$ implies $\pih^{(m)}=\pih^{(m)}_{-g}$.  Conditioning on
$P_{-A^{(m)}_g}$ for each group $g$ in
expression~\eqref{eqn:maintheorem-midproof-new-a}, we get:

\begin{align*}
\fdr^{(m)}_u & \leq \frac{\alpha^{(m)}}{G^{(m)}} \sum_{g \in \nullsm}
u^{(m)}_g w^{(m)}_g \EE{\dotfrac{ \One{P^{(m)}_g \leq f^{(m)}_g(P)}}{
    \pih^{(m)}_{-g} f^{(m)}_g(P) }}\\ &= \frac{\alpha^{(m)}}{G^{(m)}}
\sum_{g \in \nullsm} u^{(m)}_g w^{(m)}_g \EE{\EEst{\dotfrac{
      \One{P^{(m)}_g \leq f^{(m)}_g(P)}}{ \pih^{(m)}_{-g} f^{(m)}_g(P)
  }}{P_{-A^{(m)}_g}}} \\
& \stackrel{(ii)}{=} \frac{\alpha^{(m)}}{G^{(m)}} \sum_{g \in \nullsm}
u^{(m)}_g w^{(m)}_g \EE{\frac{1}{\pih^{(m)}_{-g}}\EEst{\dotfrac{
      \One{P^{(m)}_g \leq f^{(m)}_g(P)}}{ f^{(m)}_g(P)
  }}{P_{-A^{(m)}_g}}} \\
& \leq \frac{\alpha^{(m)}}{G^{(m)}} \sum_{g \in \nullsm} u^{(m)}_g
w^{(m)}_g \EE{\frac{1}{\pih^{(m)}_{-g}}},
\end{align*}
where equality (ii) holds because $\pih^{(m)}_{-g}$ is a function of
only the $p$-values outside of group $g$, i.e., of $P_{-A^{(m)}_g}$,
while the last inequality holds by \lemref{power3}(a).

Finally, observe that independence between the different groups of
partition $m$ implies that the indicator variables $\One{P^{(m)}_h
  >\lambda^{(m)}}$ are independent Bernoullis with probabilities $\geq
1-\lambda^{(m)}$ of success. Thus, as a consequence of
\lemref{power2}, we can prove that
\begin{align}
\label{eqn:storey-wbinomial-maintheorem}
 \EE{\frac{1}{\pih^{(m)}_{-g}}} ~\leq~ \frac{G^{(m)}}{\sum\limits_{h
     \in \nullsm} u^{(m)}_h w^{(m)}_h}.
\end{align}

\noindent To establish property~\eqref{eqn:storey-wbinomial-maintheorem}, let $b \defn
(1-\lambda^{(m)}),~ d \defn |\nullsm|-1$, and define
\begin{align*}
Z \defn 1+ \sum_{h \in \nullsm, h \neq g} a_h \One{P^{(m)}_h >
  \lambda^{(m)}} \text{ where } a_h &= \frac{u^{(m)}_h
  w^{(m)}_h}{|u^{(m)}\cdot w^{(m)}|_\infty}.
\end{align*}
Since $Z \leq \frac{G^{(m)} (1-\lambda^{(m)})}{|u^{(m)} \cdot
  w^{(m)}|_\infty} \pih^{(m)}_{-g}$ as the right-hand side expression
sums over more indices than the left, applying~\lemref{power2}
guarantees that
\begin{align*}
  \EE{\frac{|u^{(m)} \cdot w^{(m)}|_\infty}{G^{(m)} (1-\lambda^{(m)})
      \pih^{(m)}_{-g}}} \leq \EE{\frac1{Z}} \leq \frac{|u^{(m)}\cdot
    w^{(m)}|_\infty}{(1-\lambda^{(m)})(|u^{(m)}\cdot
    w^{(m)}|_\infty+\sum_{h\in \nullsm, h \neq g}u^{(m)}_h
    w^{(m)}_h)}.
\end{align*}
Some simple algebra then leads to
property~\eqref{eqn:storey-wbinomial-maintheorem}.


\noindent Plugging~\eqref{eqn:storey-wbinomial-maintheorem} back into
our bounds on FDR, we finally obtain
\begin{align*}
\fdr^{(m)}_u & \leq \frac{\alpha^{(m)}}{G^{(m)}} \sum_{g \in \nullsm}
u^{(m)}_g w^{(m)}_g \EE{\frac{1}{\pih^{(m)}_{-g}}}\\ &\leq
\frac{\alpha^{(m)}}{G^{(m)}} \sum_{g \in \nullsm} u^{(m)}_g
w^{(m)}_g\frac{G^{(m)}}{\sum\limits_{h \in \nullsm} u^{(m)}_h
  w^{(m)}_h} \\ &\leq \alpha^{(m)}.
\end{align*}

\noindent \paragraph{Theorem statement (b)} The proof of statement (b)
follows the same steps as for (a), but without the need to condition
on $P_{-A^{(m)}_g}$, since we do not use adaptivity.  Define the
function $f^{(m)}_{g}(P) = \frac{w^{(m)}_g \alpha^{(m)} \kh^{(m)} }{
  G^{(m)}}$.  Then $f^{(m)}_{g}$ is a nonincreasing function of $P$,
since $\kh^{(m)}$ is a nonincreasing function of $P$.

Returning to~\eqref{eqn:maintheorem-midproof-new}, as in the proof of
statement (a), we calculate
\begin{align*}
\fdr^{(m)}_u \leq \frac{\alpha^{(m)}}{G^{(m)}} \sum_{g \in \nullsm}
u^{(m)}_g w^{(m)}_g \EE{\dotfrac{ \One{P^{(m)}_g \leq f^{(m)}_g(P)}}{
    f^{(m)}_g(P) }}.
\end{align*}
By \lemref{power3}(b), we know that $\EE{\dotfrac{ \One{P^{(m)}_g \leq
      f^{(m)}_g(P)}}{ f^{(m)}_g(P) }}\leq 1$, and therefore
\begin{align*}
  \fdr^{(m)}_u \leq \frac{\alpha^{(m)}}{G^{(m)}} \sum_{g \in \nullsm}
  u^{(m)}_g w^{(m)}_g,
\end{align*}
as claimed.

\noindent \paragraph{Theorem statement (c)} We now turn to proving the
method under reshaping. Define $f^{(m)}_g(P) = \kh^{(m)}$, and define
constant $c^{(m)}_g = \frac{w^{(m)}_g \alpha^{(m)}}{G^{(m)}}$.
Returning to~\eqref{eqn:maintheorem-midproof-new}, as before, we
calculate
\begin{align*}
\fdr^{(m)}_u \leq \frac{\alpha^{(m)}}{G^{(m)}} \sum_{g \in \nullsm}
u^{(m)}_g w^{(m)}_g \EE{\dotfrac{ \One{P^{(m)}_g \leq c^{(m)}_g \cdot
      \beta^{(m)}\big(f^{(m)}_g(P)\big)}}{ c^{(m)}_g \cdot
    f^{(m)}_g(P) }}.
\end{align*}
By \lemref{power3}(c), we know that $\EE{\dotfrac{ \One{P^{(m)}_g \leq
      c^{(m)}_g \cdot \beta^{(m)}\big(f^{(m)}_g(P)\big)}}{ c^{(m)}_g
    \cdot f^{(m)}_g(P) }}\leq 1$ since $P^{(m)}_g$ is assumed to be
super-uniform for any null group $g\in\nullsm$. Therefore,
\begin{align*}
\fdr^{(m)}_u \leq \frac{\alpha^{(m)}}{G^{(m)}} \sum_{g \in \nullsm}
u^{(m)}_g w^{(m)}_g.
\end{align*}

\noindent \paragraph{Theorem statement (d)} The proof of part (d)
combines the calculations of part (a) (where adaptivity is used) with
part (c) (where reshaping is used).  Define $f^{(m)}_g = \kh^{(m)}$
and $c^{(m)}_g = \frac{w^{(m)}_g \alpha^{(m)}}{\pih^{(m)}_{-g}
  G^{(m)}}$, where $\pih^{(m)}_{-g}$ is defined as in
equation~\eqref{eqn:pihatm_g} from part (a). Note that $c^{(m)}_g$ is
no longer a constant, but nonetheless, proceeding as in part (a), we
can calculate
\begin{align*}
\fdr^{(m)}_u &\leq \frac{\alpha^{(m)}}{G^{(m)}} \sum_{g \in \nullsm}
u^{(m)}_g w^{(m)}_g \EE{\dotfrac{ \One{P^{(m)}_g \leq c^{(m)}_g\cdot
      \beta^{(m)}\big( f^{(m)}_g(P)\big)}}{ \pih^{(m)}_{-g}\cdot
    c^{(m)}_g \cdot f^{(m)}_g(P) }}.
\end{align*}
Next we condition on the $p$-values outside the group $A^{(m)}_g$:
\begin{align*}
\fdr^{(m)}_u &\leq \frac{\alpha^{(m)}}{G^{(m)}} \sum_{g \in \nullsm}
u^{(m)}_g w^{(m)}_g \EE{\EEst{\dotfrac{ \One{P^{(m)}_g \leq
        c^{(m)}_g\cdot \beta^{(m)}\big( f^{(m)}_g(P)\big)}}{
      \pih^{(m)}_{-g}\cdot c^{(m)}_g \cdot f^{(m)}_g(P)
  }}{P_{-A^{(m)}_g}}} \\
& = \frac{\alpha^{(m)}}{G^{(m)}} \sum_{g \in \nullsm} u^{(m)}_g
w^{(m)}_g \EE{\frac{1}{\pih^{(m)}_{-g}}\cdot \EEst{\dotfrac{
      \One{P^{(m)}_g \leq c^{(m)}_g\cdot \beta^{(m)}\big(
        f^{(m)}_g(P)\big)}}{ c^{(m)}_g \cdot f^{(m)}_g(P)
  }}{P_{-A^{(m)}_g}}},
\end{align*}
where the last step holds since $\pih^{(m)}_{-g}$ is a function of
$P_{-A^{(m)}_g}$.

Finally, we apply \lemref{power3}(c) to show that each of these
conditional expected values is $\leq 1$.  Of course, the subtlety here
is that we must condition on $P_{-A^{(m)}_g}$. To do so, note that,
after fixing $P_{-A^{(m)}_g}$, the function $f^{(m)}_g(P)$ can be
regarded as a function of only the remaining unknowns (i.e., of
$P_{A^{(m)}_g}$), and is still nonwincreasing, the value $c^{(m)}_g$
is now a constant; and $P^{(m)}_g = T^m_g(P_{A^{(m)}_g})$ is indeed
super-uniform since, due to the independence of $P_{A^{(m)}_g}$ from
$P_{-A^{(m)}_g}$, its distribution has not changed. Therefore, we can
apply \lemref{power3}(c) (with the random vector $P_{A^{(m)}_g}$ in
place of $P$, while $P_{-A^{(m)}_g}$ is treated as constant), to see
that $\EEst{\dotfrac{ \One{P^{(m)}_g \leq c^{(m)}_g\cdot
      \beta^{(m)}\big( f^{(m)}_g(P)\big)}}{ c^{(m)}_g \cdot
    f^{(m)}_g(P) }}{P_{-A^{(m)}_g}}~\leq~1$, and therefore,
\begin{align*}
  \fdr^{(m)}_u \leq \frac{\alpha^{(m)}}{G^{(m)}} \sum_{g \in \nullsm}
  u^{(m)}_g w^{(m)}_g \EE{\frac{1}{\pih^{(m)}_{-g}}}.
\end{align*}
Finally, we need to bound $\pih^{(m)}_{-g}$. As in the proof of part
(a), we see that the indicator variables $\One{P^{(m)}_h
  >\lambda^{(m)}}$ are independent, since
$P^{(m)}_h=T^{(m)}_h(P_{A^m_h})$, and the sets of $p$-values $P_{A^m_h}$
are assumed to be independent from each other.  Furthermore, since
$T^{(m)}_h(P_{A^m_h})$ is assumed to be a valid $p$-value, i.e.,
super-uniform for any $h\in\nullsm$, this means that the variable
$\One{P^{(m)}_h >\lambda^{(m)}}$ is Bernoulli with chance $\geq
1-\lambda^{(m)}$ of success. Therefore, the
bound~\eqref{eqn:storey-wbinomial-maintheorem} calculated in the proof
of part (a) holds here as well, and so
\begin{align*}
\fdr^{(m)}_u \leq \frac{\alpha^{(m)}}{G^{(m)}} \sum_{g \in \nullsm}
u^{(m)}_g w^{(m)}_g \frac{G^{(m)}}{\sum\limits_{h \in \nullsm}
  u^{(m)}_h w^{(m)}_h} = \alpha^{(m)}.
\end{align*}

\noindent This concludes the proof of all four parts of \thmref{pf+}.


\section{Proofs of  supporting lemmas}
\label{sec:lemmas-proofs}

In this section, we collect the proofs of some supporting lemmas.

\subsection{Proof of super-uniformity Lemma~\lowercase{\ref{lem:power}}}
\label{sec:power-proof}

\lemref{power} follows directly from earlier work \citep{blanchard2008two,barber2016p}.  Statement (a) with
inequality (but not with equality) follows as a special case of (b),
since independence is a special case of positive dependence, and the
distribution of a null $P_i$ does not change on conditioning on an
independent set of $p$-values. Statement (c) was proved also by
\citet{blanchard2008two}.  We now prove the statements (a), (d).

\paragraph{Statement (a)} We prove the first part of \lemref{power},
under the assumptions that the function $P \mapsto f(P)$ satisfies the
leave-one-out property with respect to index $i$, and that $P_i$ is
uniformly distributed and is independent of the remaining
$p$-values. Since $\PP{P_i = 0}=0$, we ignore this possibility in
the following calculations.  Since $f$ satisfies the LOOP
condition, we have
\begin{align*}
\dotfrac{\One{P_i\leq f(P)}}{f(P)} = \frac{\One{P_i\leq
    f(\ITOZERO{P})}}{f(\ITOZERO{P})}.
\end{align*}
This can be seen by separately considering what happens when the 
numerator on the left-hand side is zero or one.

Since $P^{-i}$ determines $f(\ITOZERO{P})$, it immediately follows
that
\begin{align*}
\EEst{\dotfrac{\One{P_i\leq f(P)}}{f(P)} }{P^{-i}} &= \EEst{
  \frac{\One{P_i\leq
      f(\ITOZERO{P})}}{f(\ITOZERO{P})}}{P^{-i}}\\ &=
\frac{\PPst{P_i\leq
    f(\ITOZERO{P})}{f(\ITOZERO{P})}}{f(\ITOZERO{P})} \\ &= 1,
 \end{align*}
where the last step follows since $f$ has range $[0,1]$, and $P_i$ is uniformly distributed
and is independent of $\ITOZERO{P}$;
therefore, we may deduce that $\PPst{P_i\leq f(\ITOZERO{P})}{f(\ITOZERO{P})} =
f(\ITOZERO{P})$.  This concludes the proof of the super-uniformity
lemma under independence and uniformity.


\paragraph{Statement (d)} 

For each $\ell=1,\dots,m$, let $\nu_{\ell}$ be a probability measure on $[0,\infty)$ chosen such that $\beta_\ell(k) = \beta_{\nu_\ell}(k) = \int_{x=0}^k x \;\mathsf{d}\nu_\ell(x)$,
as in the definition of a reshaping function. Let $X_\ell\sim \nu_\ell$ be drawn independently for each $\ell=1,\dots,m$, and 
let $\nu$ be the probability measure on $[0,\infty)$ corresponding to the distribution of $Z =  \prod_{\ell=1}^m X_\ell$.
Then
\begin{align*}
c \cdot \prod_{\ell=1}^m\beta_\ell(f_\ell(P))
&=c \cdot \prod_{\ell=1}^m \left( \int_{x_\ell=0}^{f_\ell(P)} x_\ell\;\mathsf{d}\nu_\ell(x_\ell)\right)\\
&=c \cdot \int_{x_1=0}^{\infty} \dots \int_{x_m=0}^{\infty} \left( \prod_{\ell=1}^m x_{\ell} \cdot \One{x_\ell\leq f_\ell(P)}\right)\; \mathsf{d}\nu_m(x_m)\dots \mathsf{d}\nu_1(x_1)\\
&=c \cdot\EE{ \prod_{\ell=1}^m \left(X_\ell\cdot \One{X_\ell\leq f_\ell(P)}\right)}\\
&=c \cdot\EE{Z \cdot \One{X_1\leq f_1(P),\dots,X_m\leq f_m(P)}}\\
&\leq c \cdot\EE{Z \cdot \One{Z \leq \prod_{\ell=1}^m f_\ell(P)}}\\
&=c \cdot\int_{z=0}^{ \prod_{\ell=1}^m f_\ell(P)} z\;\mathsf{d}\nu(z)
=c \cdot\beta_\nu\left( \prod_{\ell=1}^m f_\ell(P)\right).
\end{align*}
Therefore,
\[
\EE{\dotfrac{\One{P_i \leq c\cdot \prod_{\ell=1}^m\beta_\ell(f_\ell(P))}}{c\cdot \prod_{\ell=1}^m f_\ell(P)}}\\
\leq  \EE{\dotfrac{\One{P_i \leq c \cdot\beta_\nu\left( \prod_{\ell=1}^m f_\ell(P)\right)}}{c\cdot \prod_{\ell=1}^m f_\ell(P)}}\leq 1,\]
where the last step holds by Lemma~\ref{lem:power}(c).



\subsection{Proof of group super-uniformity Lemma~\lowercase{\ref{lem:power3}}}
\label{sec:power2-proof}

First, note that the proof of \lemref{power3}(c) is straightforward,
by applying \lemref{power}(c). More precisely, define an augmented
vector $P' = (P_1,\dots,P_n,T(P_{A_g}))\in[0,1]^{n+1}$, and define a
function $f'(P') \defn f(P_1,\dots,P_n) = f(P)$.  Since $T(P_{A_g})$
is assumed to be super-uniform (since $g\in\nullsg$ is a null group),
this means that $P'_{n+1} = T(P_{A_g})$ is super-uniform, i.e., index
$n+1$ is a null $p$-value, in the augmented vector of $p$-values
$P'$. Then applying \lemref{power}(c), with $P'$ and $f'$ in place of
$P$ and $f$, and with index $i=n+1$, yields the desired bound.

\lemref{power3}(a) is simply a special case of \lemref{power3}(b)
since independence is a special case of positive dependence, and
conditioning on an independent set of $p$-values $P_{-A_g}$ doesn't
change the distribution of $P_{A_g}$.

For \lemref{power3}(b), our proof strategy is to reduce this statement
into a form where \lemref{power}(b) becomes applicable.  (Note that we
cannot simply take the approach of our proof of \lemref{power3}(c),
because if we define an augmented vector of $p$-values
$P'=\big(P_1,\dots,P_n,\simes_w(P_{A_g})\big)$, we do not know if this
vector is positively dependent---specifically, whether $P'$ is PRDS on
entry $P'_{n+1}=\simes_w(P_{A_g})$.)

With this aim in mind, let $\kh_g \in \{0,\dots,n_g\}$ be the number
of discoveries made by the $\bh_w$ procedure when run on the $p$-values
within group $g$ at level $f(P)$. Then, using the connection between
the Simes test and the BH procedure, we may write
\begin{align*}
\One{P_g \leq f(P)} = \One{\kh_g > 0} = \dotfrac{\kh_g}{\kh_g} =
\dotfrac{\sum_{i \in A_g} \One{P_i \leq \frac{w_i \kh_g f(P)}{n_g}
}}{\kh_g}
\end{align*}
since for the $\bh_w$ procedure at level $f(P)$, the $i$th $p$-value
$P_i$ will be rejected if and only if $P_i\leq \frac{w_i \kh_g
  f(P)}{n_g}$.  Hence, we may conclude that
\begin{align*}
\dotfrac{\One{P_g \leq f(P)}}{f(P)} = \dotfrac{\sum_{i \in A_g}
  \One{P_i \leq \frac{w_i\kh_g f(P)}{n_g} }}{\kh_g f(P)} = \frac1{n_g}
\sum_{i \in A_g} w_i \dotfrac{\One{P_i \leq \tf_g(P)}}{\tf_g(P)},
\end{align*}
where we have defined $\tf_g(P) \defn \frac{w_i\kh_g f(P)}{n_g}$.

Taking expectations on both sides and applying \lemref{power}(b)
immediately proves \lemref{power3}(b).  (Specifically, we know that $P
\mapsto \kh_g$ is a nonincreasing function of $P$, and $P\mapsto f(P)$
is also assumed to be nonincreasing; therefore, $\tf_g$ is also
nonincreasing in $P$.)


Given that \lemref{power3}(b) is proved, the proof of
\lemref{power3}(d) follows exactly the same argument as above, except
that in the very last equation, $P_i$ is replaced by $S_i$, and
\lemref{power3}(b) is invoked in place of \lemref{power}(b).


\subsection{Proof of inverse-binomial Lemma~\lowercase{\ref{lem:power2}}}\label{sec:power3-proof}

The lower bound follows immediately
from Jensen's inequality, since $\EE{Z} = 1 + b \sum_{i=1}^d a_i$.
We split the argument for the upper bound into three cases.

\paragraph{Case 1: integer weights}  First, suppose that all the weights $a_i$ are integers, that is, $a_i\in\{0,1\}$ for all $i$.
In this case, we have $Z\sim 1 + \textnormal{Binomial}(k,b)$, where
$k$ is the number of weights $a_i$ that are equal to $1$. A simple calculation shows
that
\begin{align*}
\EE{\frac{1}{1+\textnormal{Binomial}(k,b)}} &= \sum_{z=0}^k\frac1{1+z}
\binom{k}{z} b^{z} (1-b)^{k-z} \\ 
&= \frac1{b(1+k)} \sum_{z=0}^k
\binom{k+1}{z+1} b^{z+1} (1-b)^{(k+1)-(z+1)}\\
& = \frac1{b(1+k)} \cdot\PP{\textnormal{Binomial}(k+1,b)\leq k}\\
 &\leq \frac{1}{b(1+k)} =
\frac{1}{b(1+\sum_i a_i)}.
\end{align*}


\paragraph{Case 2: one non-integer weight} Suppose that exactly one of the weights $a_i$ is a non-integer.
Without loss of generality we can take
$a_1=\dots=a_k=1$, $a_{k+1}=c$, $a_{k+2}=\dots=a_n=0$, for some
$k\in\{0,\dots,n-1\}$ and some $c\in(0,1)$.  Let $A=Z_1 + \dots +
  Z_{k+1}\sim \textnormal{Binomial}(k+1,b)$, and
  $Y=Z_{k+1}\sim\textnormal{Bernoulli}(b)$.  Note that $\PPst{Y=1}{A}
  = \frac{A}{1+k}$. Then
\begin{align*}
\EE{\frac{1}{Z}} &= \EE{\frac{1}{1+A - (1-c)Y}}\\ 
&=
\EE{\EEst{\frac{1}{1+A - (1-c)Y}}{A}}\\ 
&= \EE{\frac{1}{1+A} \cdot \PPst{Y=0}{A} +
  \frac{1}{c+A} \cdot\PPst{Y=1}{A}}\\ 
&= \EE{\frac{1}{1+A} +
  \left(\frac{1}{c+A} - \frac{1}{1+A}\right)\cdot\PPst{Y=1}{A}}\\ 
  &=
\EE{\frac{1}{1+A} + \frac{1-c}{(c+A)(1+A)}\cdot\frac{A}{1+k}}\\ 
&\leq
\EE{\frac{1}{1+A} + \frac{1-c}{(c+1+k)(1+A)}\cdot\frac{1+k}{1+k}},
\end{align*}
where the inequality holds since $\frac{A}{c+A}\leq \frac{1+k}{1+k+c}$
because $0\leq A\leq k+1$.  Simplifying, we get
\begin{small}
\begin{align*}
\EE{\frac{1}{Z}} \leq \EE{\frac{1}{1+A}}\cdot \frac{2+k}{1+k+c} \leq \frac{1}{b(2+k)}\cdot\frac{2+k}{1+k+c} = \frac{1}{b(1+k+c)} = \frac{1}{b(1+\sum_i a_i)},
\end{align*}
\end{small}
where the inequality uses the fact that
$\EE{\frac{1}{1+\textnormal{Binomial}(k+1,b)}}\leq \frac{1}{b(2+k)}$
as calculated in Case 1.

\paragraph{Case 3: general case}
Now suppose that there are at least two non-integer weights,
$0<a_i\leq a_j<1$. Let $C=\sum_{\ell\neq i,j}a_{\ell}Z_{\ell}$, then
$Z=1+C+a_iZ_i+a_jZ_j$.  Let $\alpha = \min\{a_i,1-a_j\}>0$.  Then
\begin{small}
\begin{multline*}
  \EEst{\frac{1}{Z}}{C} = b^{2}\cdot\frac{1}{1+C+a_i+a_j} +
  b(1-b)\cdot\frac{1}{1+C+a_i} + b(1-b)\cdot\frac{1}{1+C+a_j} +
  (1-b)^{2}\cdot \frac{1}{C}\\ \leq b^{2}\cdot\frac{1}{1+C+a_i+a_j} +
  b(1-b)\cdot\frac{1}{1+C+(a_i-\alpha)} +
  b(1-b)\cdot\frac{1}{1+C+(a_j+\alpha)} + (1-b)^{2}\cdot \frac{1}{C},
\end{multline*}
\end{small}
where the inequality follows from a simple calculation using the
assumption that $\alpha \leq a_i\leq a_j \leq 1- \alpha$. Now, define
a new vector of weights $\tilde{a}$ where $\tilde{a}_i = a_i - \alpha,
\tilde{a}_j = a_j + \alpha$ and $\tilde{a}_\ell = a_\ell$ if $\ell
\notin \{i,j\}$.  Defining $\widetilde{Z} = 1+\sum_\ell
\tilde{a}_\ell Z_\ell$, the above calculation proves that
$\EE{\frac{1}{Z}}\leq \EE{\frac{1}{\widetilde{Z}}}$ (by marginalizing
over $C$).

Note that $\sum_i a_i = \sum_i\tilde{a}_i$, but $\tilde{a}_i$ has (at
least) one fewer non-integer weight. Repeating this process
inductively, we see that we can reduce to the case where there is at
most one non-integer weight (i.e., Case 1 or Case 2). This proves the
lemma.

\section{Proof of propositions about \pf}\label{sec:prop-proofs}

\subsection{Proof of ``maximum-corner'' Proposition~\lowercase{\ref{prop:max}}}
\label{sec:proof_max}

For each $m$, by definition of $\kh^{(m)}$, there is some
$k^{(m)}_1,\dots,k^{(m)}_{m-1},k^{(m)}_{m+1},\dots,k^{(m)}_M$ such
that
\begin{align}
  \label{eqn:in_T}
  (k^{(m)}_1,\dots,k^{(m)}_{m-1},\kh^{(m)},k^{(m)}_{m+1},\dots,k^{(m)}_M)\in\Kset\;.
\end{align}
Thus, for each $m'\neq m$, $\kh^{(m')}\geq k^{(m)}_{m'}$ by definition
of $\kh^{(m')}$.  Then
\begin{align*}
  \Sh(k^{(m)}_1,\dots,k^{(m)}_{m-1},\kh^{(m)},k^{(m)}_{m+1},\dots,k^{(m)}_M)
  \subseteq \Sh(\kh^{(1)},\dots,\kh_{m-1},\kh^{(m)},\kh_{m+1},\dots,\kh^{(M)})\;,
\end{align*}
because $\Sh(k^{(1)},\dots,k^{(M)})$ is a nondecreasing function of
$(k^{(1)},\dots,k^{(M)})$, and this immediately implies
\[  \Shm(k^{(m)}_1,\dots,k^{(m)}_{m-1},\kh^{(m)},k^{(m)}_{m+1},\dots,k^{(m)}_M)
  \subseteq \Shm(\kh^{(1)},\dots,\kh_{m-1},\kh^{(m)},\kh_{m+1},\dots,\kh^{(M)}).\] 
Therefore, for each layer $m$,
\begin{align*}
\sum_{g \in \Shm(\kh^{(1)},\dots,\kh^{(m)})} u^{(m)}_g \geq \sum_{g \in
  \Shm(k^{(m)}_1,\dots,k^{(m)}_{m-1},\kh^{(m)},k^{(m)}_{m+1},\dots,k^{(m)}_M)}
u^{(m)}_g \geq \kh^{(m)},
\end{align*}
where the second inequality holds by observation~\eqnref{in_T}, and by definition of $\Kset$ as the set
of feasible vectors.  Since this holds
for all $m$, this proves that $(\kh^{(1)},\dots,\kh^{(M)})$ is itself a feasible
vector, and hence $(\kh^{(1)},\dots,\kh^{(M)}) \in \Kset$.



\subsection{Proof of ``halting'' Proposition~\lowercase{\ref{prop:alg}}}\label{sec:proof_alg}

First we introduce some notation: let $(k^{(1)}_{(s)},\dots,k^{(M)}_{(s)})$ be
the vector after the $s$th pass through the algorithm. We prove that
$k^{(m)}_{(s)}\geq \kh^{(m)}$ for all $m,s$, by induction.  At initialization,
$k^{(m)}_{(0)} = G^{(m)} \geq \kh^{(m)}$ for all $m$. Now suppose that
$k^{(m)}_{(s-1)}\geq \kh^{(m)}$ for all $m$; we now show that $k^{(m)}_{(s)}\geq
\kh^{(m)}$ for all $m$.

To do this, consider the $m$-th layer of the $s$-th pass through
the algorithm. Before this stage, we have vectors
$k^{(1)}_{(s)},\dots,k^{(m-1)}_{(s)},k^{(m)}_{(s-1)},k^{(m+1)}_{(s-1)},\dots,k^M_{(s-1)}$,
and we now update $k^{(m)}_{(s)}$. Applying induction also to this inner
loop, and assuming that $k^{(m')}_{(s)}\geq \kh^{(m')}$ for all
$m'=1,\dots,m-1$, we can now prove that $k^{(m)}_{(s)}\geq\kh^{(m)}$.  By
definition of the algorithm,
\begin{footnotesize}
\begin{align}\label{eqn:kms}
k^{(m)}_{(s)} = \max_{k'^{(m)} \in \{0,1,\dots,G^{(m)}\}}\left\{ k'^{(m)} : \sum\limits_{g \in
  \Shm(k^{(1)}_{(s)},\dots,k^{(m-1)}_{(s)},k'^{(m)},k^{(m+1)}_{(s-1)},\dots,k^{(M)}_{(s-1)})}
u^{(m)}_g \geq k'^{(m)} \right\}.
\end{align}
\end{footnotesize}
Since $k^{(m')}_{(s)}\geq \kh^{(m')}$ for all $m'=1,\dots,m-1$, and
$k^{(m')}_{(s-1)}\geq \kh^{(m')}$ for all $m'=m+1,\dots,M$, we have
\begin{align*}
\sum \limits_{g \in
  \Shm(k^{(1)}_{(s)},\dots,k^{(m-1)}_{(s)},\kh^{(m)},k^{(m+1)}_{(s-1)},\dots,k^{(M)}_{(s-1)})}
u^{(m)}_g ~\geq~ \sum\limits_{g \in
  \Shm(\kh^{(1)},\dots,\kh^{(m-1)},\kh^{(m)},\kh^{(m+1)},\dots,\kh^{(M)})} u_g^{(m)},
\end{align*}
since $\Shm(\vec{k})$ is a nondecreasing function of $\vec{k}$ by
definition. The right-hand side of this expression is in turn $\geq
\kh^{(m)}$ by definition of $(\kh^{(1)},\dots,\kh^{(M)})$ being feasible. Therefore, $\kh^{(m)}$ is in the feasible set for
Eq.~\eqnref{kms}, and so we must have $k^{(m)}_{(s)}\geq \kh^{(m)}$. By
induction, this is then true for all $s,m$, as desired.

Now suppose that the algorithm stabilizes at
$(k^{(s)}_1,\dots,k^{(s)}_M)$, after $s$ full passes.  After
completing the $m$th layer of the last pass through the algorithm, we
had vectors
$k^{(1)}_{(s)},\dots,k^{(m)}_{(s)},k^{(m+1)}_{(s-1)},\dots,k^{(M)}_{(s-1)}$;
however, since the algorithm stops after the $s$th pass, this means
that $k^{(m')}_{(s-1)}=k^{(m')}_{(s)}$ for all $m'$.  Using this
observation in the definition of $k^{(m)}_{(s)}$, we see that
\begin{align*}
\sum \limits_{g \in
  \Shm(k^{(1)}_{(s)},\dots,k^{m-1}_{(s)},k^{(m)}_{(s)},k^{(m+1)}_{(s)},\dots,k^{(M)}_{(s)})}
u^{(m)}_g \geq k^{(m)}_{(s)}.
\end{align*}
This means that $(k_{(s)}^{(1)},\dots,k_{(s)}^{(M)})\in\Kset$, and so
$k_{(s)}^{(m)}\leq \kh^{(m)}$ for all $m$ by the definition of $\kh^{(1)},\ldots,\kh^{(m)}$ and \propref{max}.  But by the induction above,
we also know that $k_{(s)}^{(m)}\geq \kh^{(m)}$ for all $m,s$, thus completing the proof.

\section{Discussion and extensions}
\label{sec:disc}

The procedures that we have analyzed and generalized do not fully
cover the huge literature on FDR-controlling procedures. For example,
\texttt{p-filter} is a generalized multi-dimensional step-up procedure, 
but much work has
also been done on alternative styles of procedures, such as step-down,
step-up-down and multi-step methods.  For example,
\citet{benjamini1999step} propose step-down procedures that control
FDR under independence. Also, procedures by
\citet{benjamini1999distribution} and \citet{romano2006stepdown}
provably control FDR under arbitrary dependence, with
\citet{gavrilov2009adaptive} extending them to adaptive control under
independence. Two-step adaptive procedures have been analyzed
in \citet{benjamini2006adaptive} under independence, and by
\citet{blanchard2009adaptive} under dependence. Different methods of incorporating
weights into such procedures have also been studied, cf.~a different notion of the
weighted Simes $p$-value proposed by~\citet{BH97}.

The super-uniformity lemmas (Lemma~\ref{lem:power}
and, in the grouped setting, Lemma~\ref{lem:power3}),
can be used to quickly prove FDR control under dependence for many of the above procedures,
and may be a useful tool for designing new multiple testing
procedures in broader settings\footnote{An analog of the super-uniformity lemma has also been discovered in the online FDR setting and has proved useful for designing new algorithms \cite{javanmard2018online,RYWJ17,ramdas2018saffron}.}. For example, it has been used to derive a decentralized procedure for FDR control on sensor networks \cite{ramdas2017qute} and a sequential algorithm for FDR control on directed acyclic graphs \cite{ramdas2018dagger}. 
This lemma was also used to derive a ``post-selection BH procedure'' \cite{brzyski2017controlling}: if a set $S \subseteq [n]$ of hypotheses was selected by the user in an arbitrary monotone data-dependent manner (see footnote~1), one way to find a subset $T \subseteq S$ that controls the FDR is to run BH on $S$ at level $\widetilde \alpha:= \alpha |S| / n$.
Indeed,
\[
\small
\fdr = \EE{\dotfrac{\sum_{i \in S \cap \nulls} \One{P_i \leq \widetilde \alpha\frac{ | T|}{| S|} }}{| T|}} 
\leq \sum_{i \in \nulls} \frac{\alpha}{n} \cdot \EE{\dotfrac{ \One{P_i \leq \alpha\frac{ | T|}{n}} }{\alpha\frac{|T|}{n}}} \leq \alpha \frac{|\nulls|}{n}.
\]
Notice that the post-selection BH procedure reduces to BH in the absence of selection, that is when $S = [n]$.
As another particularly simple but striking example, 
consider the following novel ``structured BH procedure''\footnote{This procedure was independently 
discovered recently by \citet{katsevich2018controlling}, along with several other substantial extensions.}. Suppose we
wish to insist that only certain subsets of $[n]$ are allowed to be rejected; 
let $\mathcal{K} \subseteq 2^{[n]}$ be the set of such allowed rejection sets 
(these could be determined by known logical constraints or structural requirements).
Then, if the $p$-values are positively dependent, we may reject the largest set $T \in \mathcal{K}$ such that all its $p$-values are less than $\alpha|T|/n$. Completely equivalently, one can define $\fdphat(S) = \frac{n \cdot \max_{i \in S} P_i}{|S|}$ and reject the largest set $T \in \mathcal{K}$ such that $\fdphat(T) \leq~\alpha$. This procedure controls FDR under positive dependence due to a trivial one-line proof using Lemma~\ref{lem:power}: 
\[
\small
\fdr = \EE{\dotfrac{\sum_{i \in T \cap \nulls} \One{P_i \leq \alpha \frac{ |T|}{n} }}{|T|}} 
\leq \sum_{i \in \nulls} \frac{\alpha}{n} \cdot \EE{\dotfrac{ \One{P_i \leq \alpha \frac{|T|}{n}} }{\alpha \frac{|T|}{n}}} \leq \alpha \frac{|\nulls|}{n}.
\]

Again, notice that the structured BH procedure reduces to BH in the absence of structural constraints, that is when $\mathcal{K}=2^{[n]}$. Of course, except for special structured settings, the \emph{largest} set $T \in \mathcal{K}$ may not be efficiently computable in general. When it is infeasible, \emph{any} set $T\in \mathcal{K}$ such that $\fdphat(T) \leq \alpha$ may be chosen, and FDR control will be maintained, and heuristics can be used to find large sets. (In both the above examples, one may instead use reshaping to control for arbitrary dependence.)

While there exist works that can incorporate a single layer of groups
\cite{hu2010false}, these often provide guarantees only for the finest
partition.  Alternative error metrics have been discussed by
\citet{benjamini2014selective}, who devise a way to take a single
partition of groups into account and control a \emph{selective} FDR. This
idea has been extended by \citet{peterson2016many} and
\citet{bogomolov2017testing} to partitions that form a hierarchy
(i.e., a tree). However, none of these aforementioned papers have been extended to
handle arbitrary non-hierarchical partitions, leftover or overlapping
groups, both sets of weights, adaptivity or reshaping.
Recently, \citet{katsevich17mkf} derived a knockoff \pf~that extended
the work of \citet{barber2016p} in two ways: it allows the group
$p$-values to be formed by procedures other than Simes' (like this
paper), and it can use knockoff statistics instead of $p$-values.  In
both settings it provides FDR control at a constant (between 1 and 2)
times the target FDR. While their work can handle arbitrary
non-hierarchical partitions (since it uses the same \pf~framework)
along with knockoff statistics, it also does not handle null-proportion adaptivity, both
sets of weights, reshaping, leftover or overlapping groups, and so
on. We believe that many of the algorithmic ideas and proof techniques
(especially the lemmas) introduced here may generalize to these
related works, and could be an avenue for future work.

Finally, as a last very general extension, it was recently noted by \citet{katsevich2018towards} that the $\pf$~framework can arbitrarily ``stack'' together different layers, where each layer uses a different type of FDR-controlling algorithm (ordered testing, knockoffs, online algorithms, interactive algorithms, and so on), and the $\pf$~framework can be simply used as a wrapper to ensure internal consistency.




\section*{Acknowledgments}
We thank Wenyu Chen for helping implement the new $\pf$ algorithm.
We thank Eugene Katsevich, Etienne Roquain, 
Aditya Guntuboyina and Fanny Yang for relevant discussions.
 The authors are also thankful to audience members at the Statistics
departments of Stanford, Wharton, UC Davis, the St. Louis Workshop on
Higher Order Asymptotics and Post-Selection Inference, and the NIPS
Workshop on Adaptive Data Analysis, whose questions partly shaped this work.  
This work was supported in part
by the Office of Naval Research under grant number W911NF-16-1-0368,
 the Air Force Office of Scientific Resesarch under grant number
AFOSR-FA9550-14-1-0016, by NSF award DMS-1654076, and by an Alfred P. Sloan fellowship.

{ \bibliography{FDR} }

\begin{thebibliography}{47}
\providecommand{\natexlab}[1]{#1}
\providecommand{\url}[1]{\texttt{#1}}
\expandafter\ifx\csname urlstyle\endcsname\relax
  \providecommand{\doi}[1]{doi: #1}\else
  \providecommand{\doi}{doi: \begingroup \urlstyle{rm}\Url}\fi

\bibitem[Barber and Ramdas(2016)]{barber2016p}
Rina~Foygel Barber and Aaditya Ramdas.
\newblock The p-filter: multilayer false discovery rate control for grouped
  hypotheses.
\newblock \emph{Journal of the Royal Statistical Society: Series B (Statistical
  Methodology)}, 2016.

\bibitem[Benjamini and Bogomolov(2014)]{benjamini2014selective}
Yoav Benjamini and Marina Bogomolov.
\newblock Selective inference on multiple families of hypotheses.
\newblock \emph{Journal of the Royal Statistical Society: Series B (Statistical
  Methodology)}, 76\penalty0 (1):\penalty0 297--318, 2014.

\bibitem[Benjamini and Hochberg(1995)]{BH95}
Yoav Benjamini and Yosef Hochberg.
\newblock Controlling the false discovery rate: a practical and powerful
  approach to multiple testing.
\newblock \emph{Journal of the Royal Statistical Society: Series B (Statistical
  Methodology)}, 57\penalty0 (1):\penalty0 289--300, 1995.

\bibitem[Benjamini and Hochberg(1997)]{BH97}
Yoav Benjamini and Yosef Hochberg.
\newblock Multiple hypotheses testing with weights.
\newblock \emph{Scandinavian Journal of Statistics}, 24\penalty0 (3):\penalty0
  407--418, 1997.

\bibitem[Benjamini and Hochberg(2000)]{benjamini2000adaptive}
Yoav Benjamini and Yosef Hochberg.
\newblock On the adaptive control of the false discovery rate in multiple
  testing with independent statistics.
\newblock \emph{Journal of Educational and Behavioral Statistics}, 25\penalty0
  (1):\penalty0 60--83, 2000.

\bibitem[Benjamini and Liu(1999{\natexlab{a}})]{benjamini1999distribution}
Yoav Benjamini and Wei Liu.
\newblock A distribution-free multiple test procedure that controls the false
  discovery rate.
\newblock Technical report, Tel Aviv University, 1999{\natexlab{a}}.

\bibitem[Benjamini and Liu(1999{\natexlab{b}})]{benjamini1999step}
Yoav Benjamini and Wei Liu.
\newblock A step-down multiple hypotheses testing procedure that controls the
  false discovery rate under independence.
\newblock \emph{Journal of Statistical Planning and Inference}, 82\penalty0
  (1):\penalty0 163--170, 1999{\natexlab{b}}.

\bibitem[Benjamini and Yekutieli(2001)]{BY01}
Yoav Benjamini and Daniel Yekutieli.
\newblock The control of the false discovery rate in multiple testing under
  dependency.
\newblock \emph{The Annals of Statistics}, 29\penalty0 (4):\penalty0
  1165--1188, 2001.

\bibitem[Benjamini et~al.(2006)Benjamini, Krieger, and
  Yekutieli]{benjamini2006adaptive}
Yoav Benjamini, Abba Krieger, and Daniel Yekutieli.
\newblock Adaptive linear step-up procedures that control the false discovery
  rate.
\newblock \emph{Biometrika}, 93:\penalty0 491--507, 2006.

\bibitem[Blanchard and Roquain(2008)]{blanchard2008two}
Gilles Blanchard and {\'E}tienne Roquain.
\newblock Two simple sufficient conditions for {FDR} control.
\newblock \emph{Electronic Journal of Statistics}, 2:\penalty0 963--992, 2008.

\bibitem[Blanchard and Roquain(2009)]{blanchard2009adaptive}
Gilles Blanchard and {\'E}tienne Roquain.
\newblock Adaptive false discovery rate control under independence and
  dependence.
\newblock \emph{Journal of Machine Learning Research}, 10\penalty0
  (Dec):\penalty0 2837--2871, 2009.

\bibitem[Bogomolov et~al.(2017)Bogomolov, Peterson, Benjamini, and
  Sabatti]{bogomolov2017testing}
Marina Bogomolov, Christine~B Peterson, Yoav Benjamini, and Chiara Sabatti.
\newblock Testing hypotheses on a tree: new error rates and controlling
  strategies.
\newblock \emph{arXiv preprint arXiv:1705.07529}, 2017.

\bibitem[Brzyski et~al.(2017)Brzyski, Peterson, Sobczyk, Candes, Bogdan, and
  Sabatti]{brzyski2017controlling}
Damian Brzyski, Christine~B Peterson, Piotr Sobczyk, Emmanuel~J Candes,
  Malgorzata Bogdan, and Chiara Sabatti.
\newblock Controlling the rate of gwas false discoveries.
\newblock \emph{Genetics}, 205\penalty0 (1):\penalty0 61--75, 2017.

\bibitem[Gabriel(1969)]{gabriel1969simultaneous}
Ruben Gabriel.
\newblock Simultaneous test procedures--some theory of multiple comparisons.
\newblock \emph{The Annals of Mathematical Statistics}, pages 224--250, 1969.

\bibitem[Gavrilov et~al.(2009)Gavrilov, Benjamini, and
  Sarkar]{gavrilov2009adaptive}
Yulia Gavrilov, Yoav Benjamini, and Sanat~K Sarkar.
\newblock An adaptive step-down procedure with proven {FDR} control under
  independence.
\newblock \emph{The Annals of Statistics}, 37\penalty0 (2):\penalty0 619--629,
  2009.

\bibitem[Genovese et~al.(2006)Genovese, Roeder, and
  Wasserman]{genovese2006false}
Christopher Genovese, Kathryn Roeder, and Larry Wasserman.
\newblock False discovery control with p-value weighting.
\newblock \emph{Biometrika}, 93\penalty0 (3):\penalty0 509--524, 2006.

\bibitem[Heard and Rubin-Delanchy(2018)]{heard2018choosing}
Nicholas~A Heard and Patrick Rubin-Delanchy.
\newblock Choosing between methods of combining-values.
\newblock \emph{Biometrika}, 105\penalty0 (1):\penalty0 239--246, 2018.

\bibitem[Hochberg and Benjamini(1990)]{hochberg1990more}
Yosef Hochberg and Yoav Benjamini.
\newblock More powerful procedures for multiple significance testing.
\newblock \emph{Statistics in medicine}, 9\penalty0 (7):\penalty0 811--818,
  1990.

\bibitem[Hochberg and Liberman(1994)]{HL94}
Yosef Hochberg and Uri Liberman.
\newblock An extended {S}imes' test.
\newblock \emph{Statistics \& Probability Letters}, 21\penalty0 (2):\penalty0
  101--105, 1994.

\bibitem[Hommel(1983)]{hommel1983tests}
G~Hommel.
\newblock Tests of the overall hypothesis for arbitrary dependence structures.
\newblock \emph{Biometrische Zeitschrift}, 25\penalty0 (5):\penalty0 423--430,
  1983.

\bibitem[Hu et~al.(2010)Hu, Zhao, and Zhou]{hu2010false}
James Hu, Hongyu Zhao, and Harrison Zhou.
\newblock False discovery rate control with groups.
\newblock \emph{Journal of the American Statistical Association}, 105\penalty0
  (491), 2010.

\bibitem[Javanmard and Montanari(2018)]{javanmard2018online}
Adel Javanmard and Andrea Montanari.
\newblock Online rules for control of false discovery rate and false discovery
  exceedance.
\newblock \emph{The Annals of Statistics}, 46\penalty0 (2):\penalty0 526--554,
  2018.

\bibitem[Katsevich and Ramdas(2018)]{katsevich2018towards}
Eugene Katsevich and Aaditya Ramdas.
\newblock Towards" simultaneous selective inference": post-hoc bounds on the
  false discovery proportion.
\newblock \emph{arXiv preprint arXiv:1803.06790}, 2018.

\bibitem[Katsevich and Sabatti(2017)]{katsevich17mkf}
Eugene Katsevich and Chiara Sabatti.
\newblock Multilayer knockoff filter: Controlled variable selection at multiple
  resolutions.
\newblock \emph{arXiv preprint arXiv:1706.09375}, 2017.

\bibitem[Katsevich et~al.(2018)Katsevich, Sabatti, and
  Bogomolov]{katsevich2018controlling}
Eugene Katsevich, Chiara Sabatti, and Marina Bogomolov.
\newblock Controlling {FDR} while highlighting distinct discoveries.
\newblock \emph{arXiv preprint arXiv:1809.01792}, 2018.

\bibitem[Lehmann(1966)]{lehmann1966some}
Erich~L Lehmann.
\newblock Some concepts of dependence.
\newblock \emph{The Annals of Mathematical Statistics}, pages 1137--1153, 1966.

\bibitem[Peterson et~al.(2016)Peterson, Bogomolov, Benjamini, and
  Sabatti]{peterson2016many}
Christine~B Peterson, Marina Bogomolov, Yoav Benjamini, and Chiara Sabatti.
\newblock Many phenotypes without many false discoveries: error controlling
  strategies for multitrait association studies.
\newblock \emph{Genetic epidemiology}, 40\penalty0 (1):\penalty0 45--56, 2016.

\bibitem[Ramdas et~al.(2017{\natexlab{a}})Ramdas, Chen, Wainwright, and
  Jordan]{ramdas2017qute}
Aaditya Ramdas, Jianbo Chen, Martin~J Wainwright, and Michael~I Jordan.
\newblock Qu{TE}: Decentralized multiple testing on sensor networks with false
  discovery rate control.
\newblock In \emph{Decision and Control (CDC), 2017 IEEE 56th Annual Conference
  on}, pages 6415--6421. IEEE, 2017{\natexlab{a}}.

\bibitem[Ramdas et~al.(2017{\natexlab{b}})Ramdas, Yang, Wainwright, and
  Jordan]{RYWJ17}
Aaditya Ramdas, Fanny Yang, Martin~J Wainwright, and Michael~I Jordan.
\newblock Online control of the false discovery rate with decaying memory.
\newblock In \emph{Advances In Neural Information Processing Systems}, pages
  5655--5664, 2017{\natexlab{b}}.

\bibitem[Ramdas et~al.(2018{\natexlab{a}})Ramdas, Chen, Wainwright, and
  Jordan]{ramdas2018dagger}
Aaditya Ramdas, Jianbo Chen, Martin~J Wainwright, and Michael~I Jordan.
\newblock {DAGGER}: A sequential algorithm for {FDR} control on {DAG}s.
\newblock \emph{Biometrika}, 2018{\natexlab{a}}.

\bibitem[Ramdas et~al.(2018{\natexlab{b}})Ramdas, Zrnic, Wainwright, and
  Jordan]{ramdas2018saffron}
Aaditya Ramdas, Tijana Zrnic, Martin~J Wainwright, and Michael~I Jordan.
\newblock {SAFFRON}: an adaptive algorithm for online control of the false
  discovery rate.
\newblock In \emph{Proceedings of the 35th International Conference on Machine
  Learning}, pages 4286--4294, 2018{\natexlab{b}}.

\bibitem[Romano and Shaikh(2006)]{romano2006stepdown}
Joseph Romano and Azeem Shaikh.
\newblock On stepdown control of the false discovery proportion.
\newblock In \emph{Optimality}, pages 33--50. Institute of Mathematical
  Statistics, 2006.

\bibitem[Romano et~al.(2011)Romano, Shaikh, and Wolf]{romano2011consonance}
Joseph Romano, Azeem Shaikh, and Michael Wolf.
\newblock Consonance and the closure method in multiple testing.
\newblock \emph{The International Journal of Biostatistics}, 7\penalty0
  (1):\penalty0 1--25, 2011.

\bibitem[Sarkar(2008{\natexlab{a}})]{sarkar2008methods}
Sanat Sarkar.
\newblock On methods controlling the false discovery rate.
\newblock \emph{Sankhy{\=a}: The Indian Journal of Statistics, Series A}, pages
  135--168, 2008{\natexlab{a}}.

\bibitem[Sarkar(2008{\natexlab{b}})]{sarkar2008two}
Sanat Sarkar.
\newblock Two-stage stepup procedures controlling {FDR}.
\newblock \emph{Journal of Statistical Planning and Inference}, 138\penalty0
  (4):\penalty0 1072--1084, 2008{\natexlab{b}}.

\bibitem[Sarkar(1969)]{sarkar1969some}
Tapas~K Sarkar.
\newblock Some lower bounds of reliability.
\newblock Technical report, Stanford University, 1969.

\bibitem[Seeger(1968)]{seeger1968note}
Paul Seeger.
\newblock A note on a method for the analysis of significances en masse.
\newblock \emph{Technometrics}, 10\penalty0 (3):\penalty0 586--593, 1968.

\bibitem[Simes(1986)]{simes1986improved}
John Simes.
\newblock An improved bonferroni procedure for multiple tests of significance.
\newblock \emph{Biometrika}, 73\penalty0 (3):\penalty0 751--754, 1986.

\bibitem[Sonnemann(1982)]{sonnemann1982allgemeine}
Eckart Sonnemann.
\newblock \emph{Allgemeine L{\"o}sungen multipler Testprobleme}.
\newblock Universit{\"a}t Bern. Institut f{\"u}r Mathematische Statistik und
  Versicherungslehre, 1982.

\bibitem[Sonnemann(2008)]{sonnemann2008general}
Eckart Sonnemann.
\newblock General solutions to multiple testing problems.
\newblock \emph{Biometrical Journal: Journal of Mathematical Methods in
  Biosciences}, 50\penalty0 (5):\penalty0 641--656, 2008.

\bibitem[Sonnemann and Finner(1988)]{sonnemann1988vollstandigkeitssatze}
Eckart Sonnemann and Helmut Finner.
\newblock Vollst{\"a}ndigkeitss{\"a}tze f{\"u}r multiple testprobleme.
\newblock In \emph{Multiple Hypothesenpr{\"u}fung/Multiple Hypotheses Testing},
  pages 121--135. Springer, 1988.

\bibitem[Storey(2002)]{Storey02}
John Storey.
\newblock A direct approach to false discovery rates.
\newblock \emph{Journal of the Royal Statistical Society: Series B (Statistical
  Methodology)}, 64\penalty0 (3):\penalty0 479--498, 2002.

\bibitem[Storey et~al.(2004)Storey, Taylor, and Siegmund]{Storey04}
John Storey, Jonathan Taylor, and David Siegmund.
\newblock Strong control, conservative point estimation and simultaneous
  conservative consistency of false discovery rates: a unified approach.
\newblock \emph{Journal of the Royal Statistical Society: Series B (Statistical
  Methodology)}, 66\penalty0 (1):\penalty0 187--205, 2004.

\bibitem[Stouffer et~al.(1949)Stouffer, Suchman, DeVinney, Star, and
  Williams~Jr]{stouffer1949american}
Samuel~A Stouffer, Edward~A Suchman, Leland~C DeVinney, Shirley~A Star, and
  Robin~M Williams~Jr.
\newblock The {A}merican soldier: Adjustment during army life (studies in
  social psychology in {W}orld {W}ar {II}), vol. 1.
\newblock 1949.

\bibitem[Tukey(1953)]{tukey1953problem}
John Tukey.
\newblock \emph{The Problem of Multiple Comparisons: Introduction and Parts A,
  B, and C}.
\newblock Princeton University, 1953.

\bibitem[Tukey(1994)]{tukey1994}
John Tukey.
\newblock \emph{The Collected Works of John W. Tukey, Volume VIII: Multiple
  Comparisons, 1948-1983}.
\newblock Chapman and Hall (New York), 1994.

\bibitem[Vovk and Wang(2018)]{vovk2012combining}
Vladimir Vovk and Ruodu Wang.
\newblock Combining p-values via averaging.
\newblock \emph{arXiv preprint arXiv:1212.4966v4}, 2018.

\end{thebibliography}

\bibliographystyle{plainnat}


\appendix


\section{Generalized \texorpdfstring{$\simes$}{Lg} global null tests}
\label{app:simes}

\citet{simes1986improved} proposed an improvement to the Bonferroni
procedure for global null testing at level $\alpha$. We first
calculate the Simes p-value using a reshaping function $\tbeta$ if
required:\footnote{Here and henceforth, the tilde in $\tbeta$ is used
  to signified a reshaping function for calculating a Simes p-value
  within a single group, and we will continue the use of notation
  $\beta$, without the tilde, when comparing these p-values across
  multiple groups.}
\begin{align*}
\simes(P) = \min_{1\leq k\leq n}\frac{P_{(k)}\cdot n}{\tbeta(k)},
\end{align*}
and we reject $H_{GN}$ if $\simes(P)\leq \alpha$.  
The connection to the $\by$ procedure \cite{BY01} is quite transparent: 
note that $\simes(P) \leq \alpha$ if and only if the $\by$ procedure
makes at least one rejection at level $\alpha$.

It is well known that the Simes p-value $\simes(P)$ is a bonafide
p-value, a result to be recovered as a special case of
\propref{simesw}.

\subsection{The prior-weighted \texorpdfstring{$\simes_w$}{Lg} test for
  the global null}

The Simes test~\cite{simes1986improved} was extended by~\citet{HL94}
to incorporate prior weights under independence.  As before, we define
weighted p-values $Q_i \defn P_i/w^{(1)}_i$ for each hypothesis, and
then calculate the generalized $\simes_w$ p-value for the group as
\begin{align}
\label{eqn:wSimes}
\simes_w(P) \defn \min_{1\leq k\leq n}\frac{Q_{(k)}\cdot
  n}{k}.
\end{align}
The global null hypothesis for the group $A_g$, i.e., the hypothesis that $A_g\subseteq\nulls$ consists
entirely of nulls, is then rejected at the level $\alpha$ if $\simes_w(P)\leq
\alpha$.

In a more general setting where the individual p-values $P_i$ within
the group $A_g$ may be arbitrarily dependent, we can instead consider
the reshaped weighted Simes p-value, given by
\begin{align}
\label{eqn:rwSimes}
\rsimes_w(P) \defn \min_{1\leq k\leq n}\frac{Q_{(k)}\cdot
  n}{\tbeta(k)},
\end{align}
for a reshaping function $\tbeta$ (recall definition~\eqref{eqn:reshaping}).

The following result states that the (weighted and/or reshaped) Simes
p-value really is a bonafide p-value.
\begin{proposition}
\label{prop:simesw}
Under the global null hypothesis, the weighted Simes p-value has the
following properties:
\begin{enumerate}
\item[(a)] Under independence and uniformity, if $\max_i w_i \leq
  1/\alpha$, $\simes_w(P)$ is exactly uniformly distributed.
\item[(b)] Under positive dependence (PRDS), $\simes_w(P)$ is super-uniformly
  distributed.
\item[(c)] Under arbitrary dependence, the reshaped Simes p-value
  $\rsimes_w(P)$ is super-uniformly distributed.
\end{enumerate}
\end{proposition}
While statement (a) was first proven by~\citet{HL94}, and statement
(c) under unit weights by~\citet{hommel1983tests}, all the above
statements are straightforward consequences of the properties of the
weighted BH and BY procedures.


\section{Leave-one-out property (LOOP)}
\label{app:LOOP}

For a vector $x \in \R^n$, we use $\ITOZERO{x} ~:=~
(x_1,\dots,x_{i-1},0,x_{i+1},\dots,x_n) ~\in~ \R^n$ to denote a vector
with the $i$-th coordinate set to zero.
\begin{definition}[LOOP]
  \label{def:LOOP}
A function $f:[0,1]^n\rightarrow[0,\infty)$ is said to satisfy the
  \emph{leave-one-out property} (LOOP) if for any null index $i \in
  \nulls$ and any $P \in[0,1]^n$, we have $f(\ITOZERO{P})>0$ and
\begin{align}
\label{eqn:LOOP}
\small
\begin{cases}
\text{ if } P_i \leq f(P),\text{ then } P_i \leq f(\ITOZERO{P}) =
f(P), \\ \text{ if } P_i > f(P),\text{ then } P_i > f(\ITOZERO{P}).
\end{cases}
\end{align}
\end{definition}
When $f$ satisfies LOOP, even though threshold $f(\ITOZERO{P})$ may
differ significantly from $f(P)$, the p-value $P_i$ will either lie
below both thresholds, or above both thresholds---in other words, from
the perspective of $P_i$, the threshold might as well have been
$f(\ITOZERO{P})$ instead of $f(P)$.



\section{Properties of dotfractions}
\label{app:dotfrac}

In this section, we verify that ``dotfractions'' satisfy many of the
same properties as ordinary fractions, and thus the notation
$\dotfrac{a}{b}$ can be safely used throughout the proofs of our main
results.  In all of the following, the property will be shown to hold
assuming that all dotfractions appearing in its equation or inequality
are well defined.  Hence, throughout, we assume that the various
properties are only used if all of the dotfractions in the expression
are defined---that is, we may use these properties only if we never
have $\dotfrac{a}{b}$ with $a\neq 0$ and $b=0$.  As a side note,
observe that in the paper, we always use $\dotfrac{a}{b}$ when $a,b
\geq 0$ only.
\begin{enumerate}
\item Comparing two fractions:
\begin{align}\label{eqn:dotfrac_compare}
\text{If $a \geq b\geq  0$ and $c\geq 0$, then $\dotfrac{a}{c}\geq \dotfrac{b}{c}$, and  $\dotfrac{c}{a}\leq\dotfrac{c}{b}$.}
\end{align}
In order to prove the first bound, if $c>0$ then this reduces to
$\frac{a}{c}\geq \frac{b}{c}$, while if $c=0$ then we must have
$a=b=0$ (since, otherwise, $\dotfrac{a}{c}$ and $\dotfrac{b}{c}$ would
be undefined) and so $\dotfrac{a}{c}=\dotfrac{b}{c}=0$.  To prove the
second bound, if $b>0$, then this reduces to $\frac{c}{a}\leq
\frac{c}{b}$, while if $b=0$ then we must have $c=0$ (since,
otherwise, $\dotfrac{c}{b}$ would be undefined), in which case
$\dotfrac{c}{a}=\dotfrac{c}{b}=0$.
\item Comparing against a scalar:
\begin{align}\label{eqn:dotfrac_compare_scalar}
\text{If $c\geq 0$ and $a\geq \dotfrac{b}{c}$ then $ac\geq b$.}
\end{align}
To prove this, if $c\neq 0$ then we have $a\geq \frac{b}{c}$, while if
$c=0$ then we must have $b=0$ (so that $\dotfrac{b}{c}$ is not
undefined), and so $ac\geq b$ is trivially true as both sides equal
zero.
\item Adding numerators:
  \begin{align}
    \label{eqn:dotfrac_add}
\text{For any $a,b,c$, \quad $\dotfrac{a}{c} + \dotfrac{b}{c} =
  \dotfrac{a+b}{c}$.}
\end{align}
In order to prove this claim, we note thatif $c\neq 0$ then this
reduces to $\frac{a}{c}+\frac{b}{c}=\frac{a+b}{c}$, while if $c=0$
then we must have $a=b=0$ (otherwise the dotfractions are undefined),
and so the left-hand and right-hand sides are both equal to zero.
\item Multiplying fractions:
  \begin{align}
    \label{eqn:dotfrac_mult}
\text{For any $a,b,c,d$, \quad $\dotfrac{a}{b}\cdot \dotfrac{c}{d} =
  \dotfrac{ac}{bd}$.}
\end{align}
In order to prove this claim, if $b,d\neq 0$ then this reduces to
$\frac{a}{b}\cdot \frac{c}{d} = \frac{ac}{bd}$, while if $b=0$ or
$d=0$, then either $a=0$ or $c=0$ (otherwise $\dotfrac{a}{b}$ or
$\dotfrac{c}{d}$ would be undefined), and so the left-hand and
right-hand sides are again both equal to zero.
\item Cancelling nonzero factors :
  \begin{align}
  \label{eqn:dotfrac_mult_one}
\text{If $c\neq 0$ then for any $a,b$, \quad $\dotfrac{ac}{bc} =
  \dotfrac{a}{b}$.}
\end{align}
To see why, we simply apply~\eqnref{dotfrac_mult} with $d=c$ (noting
that, with the assumption $c\neq 0$, we have $\dotfrac{c}{c}=1$).
\item Multiplying by a scalar:
  \begin{align}
    \label{eqn:dotfrac_mult_scalar}
\text{For any $a,b,c$, \quad $c\cdot \dotfrac{a}{b} =
  \dotfrac{ac}{b}$.}
\end{align}
To see why, if $b\neq 0$ then this reduces to $c\cdot
\frac{a}{b}=\frac{ac}{b}$, while if $b=0$ then we must have $a=0$ so
that $\dotfrac{a}{b}$ is not undefined, and so the left- and
right-hand sides are both zero.
\end{enumerate}

While the above properties all carry over from fractions to
dotfractions, there are some settings where familiar manipulations
with fractions may no longer be correct. For example,
$\dotfrac{a}{b}\neq\dotfrac{ac}{bc}$ when $a,b\neq 0$ while $c=0$.
Relatedly, we cannot add fractions in the usual way,
i.e.~$\dotfrac{a}{b} +\dotfrac{c}{d}$ may not be equal to
$\dotfrac{ad+bc}{bd}$; this fails because implicitly we would be
assuming that $\dotfrac{a}{b}=\dotfrac{ad}{bd}$ and
$\dotfrac{c}{d}=\dotfrac{bc}{bd}$ in order to make the two
denominators the same, which may fail if $d=0$ or if $b=0$.

\end{document}